\documentclass[12pt]{article}
\newlength{\extraspace}
\newlength{\extraspaces}
\usepackage{graphicx}
\usepackage{latexsym}
\usepackage{amsmath}
\usepackage{amsfonts}
\usepackage{hyperref}

\renewcommand{\theequation}{\thesection.\arabic{equation}} \csname
@addtoreset\endcsname{equation}{section}

\def\cA{{\cal A}}
\def\cB{{\cal B}}

\def\cL{{\cal L}}

\def\cN{{\cal N}}
\def\cO{{\cal O}}

\def\cU{{\cal U}}
\def\cV{{\cal V}}
\def\cW{{\cal W}}

\def\be{\begin{equation}}
\def\ee{\end{equation}}
\def\bea{\begin{eqnarray}}
\def\eea{\end{eqnarray}}
\def\ba{\begin{array}}
\def\ea{\end{array}}

\begin{document}

\begin{flushright}

\end{flushright}

\vspace{40pt}

\begin{center}

{\Large\sc Exploring Three-dimensional Higher-Spin Supergravity based on $sl(N|N-1)$ Chern-Simons theories}

\vspace{50pt}

{\sc Hai Siong Tan}

\vspace{15pt}
{\sl\small
Berkeley Center for Theoretical Physics and Department of Physics,\\
	University of California, Berkeley, CA 94720-7300\\

\vspace{10pt}
{\it haisiong$\_$tan@berkeley.edu}}

\vspace{70pt} {\sc\large Abstract}\end{center}
We investigate various aspects of higher-spin anti-de Sitter supergravity in three dimensions as described by Chern-Simons theory based on the finite-dimensional superalgebra $sl(N|N-1)$, with the particular case of $N=3$ as our prime example. This class of theories serves as a natural supersymmetrization of the higher-spin gravity theory based on $sl(N)$ Chern-Simons theories. We demonstrate explicitly that the asymptotic symmetry algebra contains the $\cN=2$ superconformal algebra in each sector. The appropriate Killing spinor equations are derived and used to classify existing and new classical solutions. We also discuss holonomy conditions, higher-spin black holes and conical defect spacetimes in this class of theories.
\newpage

\tableofcontents
\section{Introduction}\label{sec:Intro}

Higher-spin gravity in various dimensions, as introduced in the seminal papers \cite{vasiliev_int} and \cite{Bekaert:2005vh}, has recently gathered much interest because, among other reasons, it appears to furnish a relatively more manageable context to understand principles of holography as compared to say, a full-fledged string theory. In four dimensions, many insights have been gathered from a conjectured duality \cite{Sezgin:2002rt, Klebanov:2002ja} between higher-spin gravity in $AdS_4$ and $O(N)$ vector models\footnote{Most recently, \cite{Chang:2012kt} proposes some string-theoretic realizations.} , which recently inspired proposals to understand a higher spin realization of dS/CFT by analytically continuing bulk and boundary correlators in the $AdS_4$ case \cite{Anninos:2011ui}. 

In three dimensions, consistent higher spin gravitational theories can be written down with only a finite number of higher spin fields \cite{Blencowe}. These massless gauge fields are higher spin counterparts of the topological graviton in three dimensions. Recall that in ordinary gravity, as first pointed out in \cite{Witten} and \cite{Achucarro:1987vz}, the Einstein-Hilbert action (with a negative cosmological constant) in three dimensions can be written as the difference between two Chern-Simons actions, each equipped with a $sl(2)$ gauge algebra\footnote{Taking the sum instead of the difference leads to topologically massive gravity \cite{Deser, Deser:1982} of which higher-spin analogue is studied in \cite{Chen, Bagchi}.}. 

The global degrees of freedom arise from the boundary excitations of the bulk fields, and a Brown-Henneaux analysis \cite{Brown} reveals left and right-moving Virasoro algebras as its asymptotic symmetry group. A natural generalization is to replace the $sl(2)$ with $sl(N)$ for some positive integer $N>2$. This yields a class of higher-spin gravity theories, with the two copies of Virasoro algebras enlarged to two copies of $\cW_N$ algebras. One can take a certain large $N$ limit to obtain $hs[\lambda] \oplus hs[\lambda]$ Chern-Simons theory with $\cW_{\infty} [\lambda]$ as the asymptotic symmetry algebra, with $\lambda$ being a deformation parameter of the infinite-dimensional higher-spin algebra. Of prime motivation to us, in this paper, is a holographic duality due to Gaberdiel and Gopakumar \cite{Gaberdiel:2010pz}, that relates this theory coupled to a complex massive scalar to a class of CFTs defined on the boundary. 

Let us now briefly summarize the salient points of this duality conjecture\footnote{We shall follow the latest paper by the two authors in \cite{Gaberdiel:2012ku} which proposes to refine some aspects of the original conjecture in \cite{Gaberdiel:2010pz}.}. In the bulk, we have an infinite tower of higher-spin fields coupled to a complex scalar of mass-squared $M^2 = -1+\lambda^2$. The dual CFT is conjectured to be a 't Hooft limit of the $\cW_N$ minimal coset model of the form 
$$
\frac{su(N)_k \oplus su(N)_1}{su(N)_{k+1}},\qquad k,N \rightarrow\infty,\,\, \lambda = \frac{N}{k+N}
$$
with the 't Hooft parameter $\lambda$ fixed. This duality conjecture relies on how the $\cW$-algebra dictates the representation theory of the minimal CFT in the above limit \cite{Gaberdiel:2011wb}, and even at finite $N$ and $k$, it is strengthened by a recently observed isomorphism between quantum $\cW_{\infty}[\lambda]$ with three distinct $\lambda$ \cite{Gaberdiel:2012ku}. Some evidence for it includes: the one-loop determinant of the gravitational sector being equal to the vacuum character of $\cW_N$ \cite{Gaberdiel:2011zw} and the matching of bulk conical defect solutions to light states of the CFT \cite{Castro:2011iw}. 

Supersymmetrization of this duality conjecture has also been most recently discussed. Let us very briefly mention some recent progress. In \cite{Creutzig:2011fe, Candu:2012jq} , the gauge algebra $shs[\lambda] \oplus shs[\lambda]$ is considered and the proposed dual CFT is the 't Hooft limit of the super-coset $\cN = (2,2)$ two-dimensional Kazama-Suzuki model. In particular, the vacuum character of the $\cN=(2,2)$ $\cW_{\infty}[\lambda]$ algebra was computed and checked to agree remarkably with the massless sector of the bulk partition function. In \cite{Henneaux:2012ny}, the authors considered quite generically $(N,M)$-extended higher-spin $AdS_3$ SUGRA and analyzed their asymptotic spacetime symmetries, paying particular attention to $shs(N|2) \oplus shs(M|2)$ gauge algebras.\footnote{In the notation of \cite{Henneaux:2012ny}, $shs(2|2)$ algebra generally refers to the undeformed $shs[\lambda]$ algebra that is referred to in this paper and \cite{Creutzig:2011fe}, i.e. taking $\lambda = \frac{1}{2}.$} They demonstrated (see also \cite{Cheng}) that the asymptotic symmetry is enhanced to some $(N,M)$-extended super-$\cW_{\infty}[\lambda]$ nonlinear superalgebra. 

In this paper, we will take some modest steps in understanding $\cN =(2,2) $ higher-spin $AdS_3$ SUGRA as described by Chern-Simons theory based on the finite-dimensional superalgebra $sl(N|N-1)\oplus sl(N|N-1)$. This class of theories serves as a natural supersymmetrization of the higher-spin gravity theory based on $sl(N)\oplus sl(N)$ theories. It has some possible relevance for a supersymmetric version of the Gaberdiel-Gopakumar conjecture, as was first pointed out recently in \cite{Creutzig:2011fe}. The supersymmetry of the higher-spin gravity theory constructed here is defined by the $osp(2|2)$ superalgebra that is present as a sub-superalgebra. We derive the appropriate Killing spinor equations, and classify both existing and new classical solutions we found, based on the number of real Killing spinors preserved. 
Another closely related development is the study of higher-spin black holes in $sl(N)$ and $hs[\lambda]$ Chern-Simons theories, as was first presented in \cite{Kraus1}, \cite{Kraus2} and \cite{Castro:2011fm}. These black holes (see also \cite{Tan:2011tj} for some details in the $sl(4)$ case) are characterized by a trivial holonomy along the Euclidean time-like direction, and they exhibit some rather remarkable integrability conditions that enable one to write down sensible black hole thermodynamical laws, and even reproduce their classical partition function from the dual CFT. In this paper, we will pay some attention to holonomy conditions and briefly discuss higher-spin black holes in the $sl(N|N-1)$ theories.

The outline of this paper is as follows: in Section \ref{sec:two}, we review the basic formulation of higher-spin gravity in the framework of Chern-Simons theory based on $sl(M|M-1)$. We review the group structure in the `Racah' basis which we find to be most convenient. In Section \ref{sec:three}, we carry out a standard analysis to recover the $\cN=2$ superconformal algebra from the asymptotic spacetime algebra, in the process, performing a Sugawara redefinition of the energy-momentum tensor. In Section \ref{sec:four}, we derive the appropriate Killing spinor equations by considering a particular gauge symmetry, and then classify these classical solutions based on the number of real Killing spinors preserved. We also construct new classical solutions which are natural supersymmetrization of the black holes and conical singularities of the corresponding $sl(N)$ theories, and briefly noted some differences with the $sl(N)$ case. Finally, in Section \ref{sec:five}, we extend our results to $sl(N|N-1)$ theories for a general finite $N$ (also briefly discussing the infinite-dimensional $shs[\lambda]$ superalgbra case). In Section \ref{sec:discussion}, we summarize the main results of our paper and suggest some future directions. 

\section{Higher-spin supergravity and Chern-Simons theory based on $sl(N|N-1)$ superalgebra}
\label{sec:two}

\subsection{General remarks on higher-spin $AdS_3$ SUGRA as a Chern-Simons theory}
\label{sec:Prelim}
Below, we review some basics of ordinary and higher-spin anti-de Sitter SUGRA based on Chern-Simons theory, mainly following the introduction in \cite{Henneaux:2012ny}. There are two basic requirements of the superalgebra associated with an $AdS_3$ SUGRA: (i) it contains an $sl(2,\mathbb{R}) \oplus sl(2,\mathbb{R})$ as a sub-algebra, and (ii) the fermionic generators transform in the \textbf{2} of the $sl(2)$. This condition is satisfied by seven classes of superalgebras (see table 1 of \cite{Henneaux:2012ny}. In this paper, our main interest lies in a higher-spin counterpart of $osp(2|2,\mathbb{R})\oplus osp(2|2,\mathbb{R})$, which belongs to the general class of $osp(N|2,\mathbb{R}) \oplus osp(M|2,\mathbb{R})$ Chern-Simons theory (these are SUGRA theories which have $\cN =(N,M)$ supersymmetry). 

In the non-supersymmetric case, a consistent higher spin gravity can be defined as a $sl(N,\mathbb{R})$ Chern-Simons theory which includes higher spin gauge fields up to spin $\leq N$. In a particular $N\rightarrow \infty$ limit, the gauge algebra becomes an infinite-dimensional $hs(2,\mathbb{R})$ algebra. Supersymmetrization of this construction was first discussed in \cite{Henneaux:2012ny} in which the infinite-dimensional superalgebra $shs(N|2,\mathbb{R})\oplus shs(M|2,\mathbb{R})$ is studied as the higher-spin extension of $osp(N|2,\mathbb{R})\oplus osp(M|2,\mathbb{R})$. The first can be represented as the quotient of the universal enveloping algebra of the latter by a certain ideal.

In the framework of Chern-Simons theory, supersymmetrization implies among other things, that the Lie algebra is replaced by suitable superalgebras, along with the supertrace in place of the ordinary one. In this paper, we will study the particular case of $sl(N|N-1)$ superalgebras as the higher-spin gauge algebras, with $N=3$ as our main example. We should remark that $sl(N|N-1)$ is not a consistent truncation of $shs[\lambda]$ in the sense that it is \emph{not} a subalgebra of the latter. Indeed, the only non-trivial sub-superalgebra of $shs[\lambda]$ is $osp(2|2)$. Nonetheless, $sl(N|N-1)$ contains $sl(2|1) \simeq osp(2|2)$ as a sub-superalgebra, and further there is a well-defined analytic continuation procedure to send it to $shs[\lambda]$. This situation appears identically in the $sl(N)$ case. As noted in \cite{Henneaux:2010xg}, $sl(3)$ cannot be obtained as an algebraic truncation of $hs[\lambda]$. However, if we force terms valued in higher spin fields of spin greater than two to be zero by hand, then the truncated algebra is isomorphic to $sl(3)$. We are interested in $sl(N|N-1)$ as a higher-spin SUGRA theory which is a natural supersymmetric generalization of the $sl(N)$ higher-spin theories.

Another important relation which we will discuss further is that to super-$\cW$ algebras. As is well-known, the Drinfeld-Sokolov Hamiltonian reduction procedure, when applied to WZW models, takes affine $sl(2)$ current algebra to Virasoro algebra. Generalization of this method has been recently used to explain how $\cW_N$ and $\cW_{\infty}[\lambda]$ can be obtained from affine $sl(N)$ and $hs[\lambda]$ algebras respectively, in the context of higher-spin gravitational theories. Supersymmetrization of this computation was done in \cite{Henneaux:2012ny}. In this paper, although we do not explicitly carry out the full computation to obtain the classical $\cN=2\,\, \cW_3$ algebra (see \cite{Romans:1991wi} for the full quantum $\cN=2\,\,\cW_3$ algebra and \cite{Lu} for the classical algebra), in Section \ref{sec:three}, we shall truncate some of the gauge fields to recover the $\cN=2$ superconformal algebra, and in the process, compute the Sugawara redefinition of the energy-momentum tensor. We should note that for non-higher spin supergravity theories, the relationship between superconformal algebras and asymptotic dynamics has been explained elegantly in the nice work of \cite{Henneaux:1999ib} for the finite-dimensional gauge algebras, and the Hamiltonian reduction procedure explained in \cite{Coussaert:1995zp}. 

In the ordinary case of $osp(N|2)$ gauge algebra, the odd-graded generators are fundamental spinors of $sl(2)$ and vectors of $so(N)$, while the even-graded ones consist of the sum of $sl(2)$ and $so(N)$ generators. 
In particular, the $N=2$ case is important for us due to the isomorphism $osp(2|2) \simeq sl(2|1)$, the latter being a sub-superalgebra of $sl(N|N-1)$.  Henceforth, we will pay attention to the $N=3$ case. The Killing spinor equations were solved for a number of classical backgrounds, and global $AdS_3$ and the massless BTZ black hole \cite{BTZ} arise as the Neveu-Schwarz and Ramond vacua of the theory. The supersymmetric higher-spin theories based on $sl(N|N-1)$ gauge algebra will have $\cN=2$ supersymmetry in either/both chiral sectors of the Chern-Simons theory, due to the $osp(2|2)$ sub-superalgebra.

\subsection{About $sl(N|N-1)$ in the Racah basis}
\label{sec:racah}
Now, the general $sl(N|N-1)$ element can be decomposed as \cite{Fradkin:1990qk}
\be
\label{groupstructure}
sl(N|N-1) = sl(2) \oplus \left( \bigoplus_{s=2}^{N-1} g^{(s)} \right) \oplus \left( \bigoplus_{s=0}^{N-2} g^{(s)} \right) \oplus 2\times \left( \bigoplus_{s=0}^{N-2} g^{(s +\frac{1}{2})}  \right)
\ee
where $g^{(s)}$ is defined as a spin-$s$ multiplet of $sl(2)$. This is one less than the conformal or spacetime spin\footnote{From now on, by `spin' we refer to the $sl(2)$-spin unless otherwise stated.}. Hence, for example, for $sl(3|2)$, the even-graded sector will consist of three $sl(2)$ generators, five spin-2 generators, one abelian gauge field, one spin-1 field, whereas the odd-graded part consists of two copies of a spin-1/2 multiplet and a spin-3/2 multiplet. 

In this paper, we take the action to be the difference between two super-Chern Simons action at level $k$,
\be
\label{CSaction}
S_{CS} [\Gamma, \tilde{\Gamma}]=\frac{k}{4\pi} \int \text{str} \left( \Gamma\wedge d\Gamma + \frac{2}{3} \Gamma\wedge \Gamma \wedge \Gamma \right) - \frac{k}{4\pi} \int \text{str} \left( \tilde{\Gamma}\wedge d\tilde{\Gamma}+ \frac{2}{3} \tilde{\Gamma}\wedge \tilde{\Gamma} \wedge \tilde{\Gamma} \right)
\ee
where `str' stands for the super-trace, and $\Gamma$(chiral sector) and $\tilde{\Gamma}$(anti-chiral sector) are the connection one-forms valued in the elements of the superalgebra. When the higher-spin theory is cast as a Chern-Simons theory, it is essential to specify how the gravitational $sl(2)$ sector is embedded in the gauge algebra. For example, in the $sl(N)$ case, Chern-Simons theory based on the gauge algebra $sl(N)$ can realize physically distinct higher-spin theories arising from inequivalent embeddings of $sl(2)$.\footnote{There can however be problems with unitarity for the non-principal embeddings as discussed in \cite{Castro:Unitarity}.} Demanding that the gravitational $sl(2)$ is part of the $osp(2|2)$ superalgebra, we can work in the so-called `Racah' basis to write down the commutation relations that are more suited for us to identify the physical interpretation of various fields. We will leave explicit details of the matrix realization of the superalgebra to the Appendix \ref{AppB}, but now, let us summarize and review some essential points about the way we describe the $sl(N|N-1)$ in the Racah basis following \cite{Fradkin:1990qk}.

It is convenient to start from $gl(N|N-1)$ and obtain $sl(N|N-1)$ by quotienting out its center. The generators are $\mathbb{Z}_2$-graded by the usual Grassmann parity function: the bosonic ones we denote by $T, U$ which generate $gl(N)$ and $gl(N-1)$ in \eqref{groupstructure}, and the fermionic ones by $Q,\bar{Q}$ which generate the half-integer spin multiplets.\footnote{Formally, let the Grassmann parity function be $P(T)=P(U)=1=-P(Q)=-P(\bar{Q})$, and define the supercommutator by $[A,B\} = AB - (-1)^{P(A)P(B)}BA$.} The supercommutation relations read schematically (the structure constants are multiples of Wigner $6j$-symbols (see Appendix \ref{AppB})):
\bea
&&[T,T]\sim T,\qquad [U,U]\sim U, \qquad \{Q,\bar{Q}\}\sim T + U \\
&&[T,\bar{Q}] \sim \bar{Q}, \qquad [U,\bar{Q}] \sim \bar{Q}, \qquad [T,Q] \sim Q, \qquad  [U,Q] \sim Q.
\eea
In the notation of \eqref{groupstructure}, we denote $T^{s}_{m}, -s\leq m\leq s$, to generate each $g^{(s)}$ multiplet, and similarly for $U,Q,\bar{Q}$. The identity matrix $\mathbf{1} = \sqrt{N} T^0_0 + \sqrt{N-1}U^0_0$ is the center, and after modding it out, we have $sl(N|N-1) \sim gl(N|N-1)/\mathbf{1}$. 

Further for our purpose, we should re-define the generators such that we can form a basis for the $sl(2)$ in \eqref{groupstructure} with all the other generators transforming under its irreducible representations. This is the gravitational $sl(2)$ sub-algebra when $sl(2)$ is embedded principally in $sl(N|N-1)$. Henceforth, we will allude to the specific case of $N=3$ as a concrete example. First, let us consider linear combinations of the $T^s_m, U^s_m$ as follows
\bea
\label{redefine}
&&L_0 = \frac{1}{\sqrt{2}} \left(2T^1_0 + U^1_0 \right), L_{\pm 1} = 2 T^1_{\pm 1} + U^1_{\pm 1}, \cr
&&A_0 = \frac{1}{\sqrt{2}} \left(2T^1_0 - U^1_0 \right), A_{\pm 1} = 2 T^1_{\pm 1} - U^1_{\pm 1}. \cr
&&W_{\pm 2} = 4T^2_{\pm 2}, \qquad W_{\pm 1} = 2T^2_{\pm 1},\qquad W_0 = \sqrt{\frac{8}{3}}T^2_0 \cr
&&U_0 = \frac{1}{\sqrt{3}} T^0_0 + \frac{1}{\sqrt{2}} U^0_0 
\eea
which realize the even part of $sl(3|2)$ as
\bea
\label{balgebra}
&&[L_i, L_j] = (i-j)L_{i+j},\qquad [A_i, A_j] = (i-j)L_{i+j},\qquad [L_i, A_j] = (i-j)A_{i+j} \cr \cr
&&[W_i, W_j] = \frac{(j-i)}{3} \left(2j^2 + 2i^2 - ij - 8 \right)\times \frac{1}{2} \left( L_{i+j} + A_{i+j} \right) \cr \cr
&&[L_i, W_j] = (2i-j) W_{i+j},\qquad [A_i, W_j]=(2i-j)W_{i+j}.
\eea
The generators $L_i$ form the basis for the total $sl(2)$. Note that the commutator between the $W$'s is the same as that in $sl(3)$ but with $A \rightarrow L$. Indeed, the subset of generators $\{ (L_i+A_i)/2, W_i\}$ forms the $sl(3)$ algebra. Also, some commutation relations involving $Q,\bar{Q}, U_0$ and $J_m$ can be written as
\be
\label{Q}
\{Q^{(\frac{1}{2})}_r, \bar{Q}^{(\frac{1}{2})}_s\} = \frac{1}{3} J^a \left(\gamma_a \gamma_0 \right)_{rs} + \left( \gamma_0 \right)_{rs} \cU_0, 
\ee
\be
\label{U}
\left[ U_0, Q^{(\frac{1}{2})}_r  \right] = \frac{1}{6} Q^{(\frac{1}{2})}_r, \qquad
\left[ U_0, \bar{Q}^{(\frac{1}{2})}_r  \right] = -\frac{1}{6} \bar{Q}^{(\frac{1}{2})}_r,
\ee
\be
\label{F}
[ J_m, Q^{(\frac{1}{2})}_r ] = -\frac{1}{2} \left( \gamma_m \right)_{rs} Q^{(\frac{1}{2})}_s, \qquad
[ J_m, Q^{(\frac{3}{2})}_r ] = \frac{1}{2}  \left( \gamma^{(\frac{3}{2})}_m \right)_{rs} Q^{(\frac{3}{2})}_s,
\ee
where $r,s = -\frac{1}{2}, \frac{1}{2}$ and we define
\bea
&& J_0 = \frac{1}{2} \left( L_1 + L_{-1} \right), \qquad  J_1 = \frac{1}{2} \left( L_1 - L_{-1} \right), \qquad  J_2 = L_0. \qquad  [J_a, J_b] = \epsilon_{abc} J^c \cr \cr
&&\gamma_0 = \begin{pmatrix}
0 & -1\\
1 & 0
\end{pmatrix},\,\,
\gamma_1 = \begin{pmatrix}
0 & 1\\
1 & 0
\end{pmatrix},\,\,
\gamma_2 = \begin{pmatrix}
-1 & 0\\
0 & 1
\end{pmatrix}. \qquad
[\gamma_a, \gamma_b] = 2\epsilon_{abc} \gamma^c.\nonumber
\eea
$$
\gamma^{(\frac{3}{2})}_0 = \left(   \begin{array}{cccc}
0 & -\sqrt{3} & 0 & 0\\
\sqrt{3} & 0 & -\sqrt{4} & 0\\
0 & \sqrt{4} & 0 & -\sqrt{3} \\
0 & 0 & \sqrt{3} & 0
\end{array}\right), \,\,\,
\gamma^{(\frac{3}{2})}_1 = -\left(   \begin{array}{cccc}
0 & \sqrt{3} & 0 & 0\\
\sqrt{3} & 0 & \sqrt{4} & 0\\
0 & \sqrt{4} & 0 & \sqrt{3} \\
0 & 0 & \sqrt{3} & 0
\end{array}\right), \,\,\,
\gamma^{(\frac{3}{2})}_2= 2[\gamma_0, \gamma_1]
$$
where $\gamma^{(\frac{3}{2})}_m$ are spin-$\frac{3}{2}$ matrix realizations of $sl(2)$. Identical relations hold for the $\bar{Q}$ generators. Thus, the fermionic generators $Q, \bar{Q}$ transform as irreducible $sl(2)$ tensors. With the identification of the gravitational $sl(2)$, we checked that the Chern-Simons level $k$ can be identified as 
\be
k=\frac{l}{4G}
\ee
provided we normalize the supertrace `str' $= \frac{1}{3} (\sum_{i=1}^{3} - \sum_{i=4}^5) M_{ii}$ where $M$ is any supermatrix.



\section{Asymptotic spacetime symmetries}
\label{sec:three}
\subsection{On super-$\cW$ symmetries as asymptotic spacetime symmetry}
\label{sec:Wsymmetry}
In the following, we shall briefly discuss the asymptotic spacetime symmetry of $sl(3|2)$ higher-spin SUGRA. This was initiated briefly in \cite{Creutzig:2011fe}. Our results generalize straightforwardly to the general $sl(N|N-1)$ cases. We begin by taking the manifold for the Chern-Simons theory (both copies) to be $\mathbb{R} \times D^2$, with co-ordinates $(\rho, \phi)$ and $t\in \mathbb{R}$. The radial co-ordinate is $\rho$ and the boundary cylinder at infinite $\rho$ is parametrized by $x^{\pm} \equiv t \pm \phi$. 

Consider the chiral sector (our analysis that follows generalizes straightforwardly to the anti-chiral sector). We can fix some of the gauge freedom by choosing
\be
\Gamma_- = 0, \qquad \Gamma_+ = e^{-\rho L_0} a(x^+) e^{\rho L_0}, \qquad \Gamma_{\rho} = L_0
\ee
where $\Gamma_{\pm} = \Gamma_t \pm \Gamma_{\phi}$. By imposing the boundary condition that we obtain an asymptotically $AdS_3$ space, one can write, in a `highest-weight' gauge,
\be
\label{gaugec}
a(x^+) = L_1 + \cL L_{-1} + \cW W_{-2} + \cU\,U_0 +  \Upsilon A_{-1} + \varphi_+ Q^{\frac{1}{2}}_{\frac{1}{2}} + \bar{\varphi}_+ \bar{Q}^{\frac{1}{2}}_{\frac{1}{2}} + \Phi_+ Q^{\frac{3}{2}}_{\frac{3}{2}} + \bar{\Phi}_+ \bar{Q}^{\frac{3}{2}}_{\frac{3}{2}}.
\ee
It can be shown straightforwardly that such a gauge choice is still preserved by gauge transformations of the form 
\be
\Lambda(x^+) = e^{-\rho L_0} \lambda (x^+) e^{\rho L_0}
\ee
where the gauge parameter $\lambda$ is valued (with $x^+$ dependence in the components suppressed in notation) in the basis generators as follows:
\bea
\lambda &=& \xi^n L_n + \chi^n  W_n + \alpha^n A_n + \eta\,U_0 + \sum_{i=-\frac{1}{2}, \frac{1}{2}} \nu_i Q^{\frac{1}{2}}_i + \sum_{i=-\frac{1}{2}, \frac{1}{2}} \sigma_i Q^{\frac{3}{2}}_i + \sum_{i=-\frac{3}{2}, \frac{3}{2}} \zeta_i Q^{\frac{3}{2}}_i \cr 
&+&\sum_{i=-\frac{1}{2}, \frac{1}{2}} \bar{\nu}_i \bar{Q}^{\frac{1}{2}}_i + \sum_{i=-\frac{1}{2}, \frac{1}{2}} \bar{\sigma}_i \bar{Q}^{\frac{3}{2}}_i + \sum_{i=-\frac{3}{2}, \frac{3}{2}} \bar{\zeta}_i \bar{Q}^{\frac{3}{2}}_i.
\eea
Performing this gauge transformation at fixed time, the gauge connection $a$ changes as
\be
\label{constraintg}
\delta a = d_{\phi}\lambda + [a, \lambda]
\ee
upon which we require that the form of the ansatz \eqref{gaugec} be preserved. This will lead to a set of constraint equations for the variations of various fields. At the boundary, the gauge transformations may generate a physically inequivalent state. These physical symmetries are generated by boundary charges. From the Chern-Simons action, one can compute the classical Poisson brackets among these charges.

In the non-supersymmetric case, this procedure yields the $W_N$ algebras\cite{Camp:2010, Camp:2011}. Essentially, imposing the condition that we have $AdS_3$ boundary conditions at infinity turns out to be equivalent to the Drinfeld-Sokolov reduction of the current algebra. In an identical fashion, we can perform the same analysis in the supersymmetric case. We solve the constraint equations due to \eqref{constraintg} and compute the variation of various fields. The various parameters are reduced to the following set of independent ones which we denote as 
\be
\eta,\,\,\, \xi\equiv \xi^{(1)},\,\,\, \chi \equiv \chi^{(2)},\,\,\,\alpha \equiv \alpha_1,\,\,\,\nu_- \equiv \nu_{-\frac{1}{2}},\,\,\,\bar{\nu}_- \equiv \bar{\nu}_{-\frac{1}{2}}, \zeta_- \equiv \zeta_{-\frac{3}{2}},\,\,\,\bar{\zeta}_- \equiv \bar{\zeta}_{-\frac{3}{2}}.
\ee
The variation of various fields can be computed straightforwardly, but it is very cumbersome even for the case of $sl(3|2)$. The OPEs can be computed straightforwardly because by the Ward identities, the fields' variations are identical to those generated via
\be
\label{noether}
\delta \cO = 2\pi \text{Res} \left( J(\phi) \cO(0)  \right)
\ee
where the Noether current $J$ taking the form of 
\be
\label{currentj}
J=\frac{1}{2\pi} \left( \xi \cL + \eta \cU + \alpha \Upsilon+ \chi \cW + \nu_- \bar{\varphi}_+ + \bar{\nu}_- \varphi_+  + \zeta_- \bar{\Phi}_+ + \bar{\zeta}_- \Phi_+ \right).
\ee
The OPEs depend on the choice of basis and are sensitive to field/parameter redefinitions. To adopt the appropriate convention, it is natural to do so such that the $\cN=2$ super-Viraoso algebra can be obtained after a truncation. We will not explicitly carry out the full computation to obtain the classical $\cN=2\,\, \cW_3$ algebra (see \cite{Lu} for the classical and \cite{Romans:1991wi} for the quantum $\cN=2\,\,\cW_3$ algebra), but in the next section, after truncating some fields, we recover the $\cN=2$ superconformal algebra, and in the process, compute explicitly the Sugawara redefinition of the energy-momentum tensor.

\subsection{Recovering the $\cN=2$ super-Virasoro algebra}
\label{sec:recover}
As explained earlier, $sl(3|2)$ contains as a sub-algebra, $sl(2|1) \simeq osp(2|2)$. Below, we demonstrate explicitly, as a consistency check, that restricting to fields valued in this subalgebra, we recover precisely the OPE relations pertaining to the well-known $\cN=2$ super-Virasoro algebra. This is the symmetry algebra that dictates the boundary degrees of freedom correponding to bulk Chern-Simons fields valued in the $sl(2|1)$ sector, as we shall shortly verify. 

In the following, we first present the relevant variations of the relevant fields $\varphi_+, \bar{\varphi}_+, \cL$ and  $\cU$. 
\bea
\label{11}
&&\delta \varphi_+ = \frac{3}{2} \varphi_+ \xi'\, + \varphi_+' \xi \, + \frac{1}{6}\cU \varphi_+ \xi - \frac{1}{6}\varphi_+ \eta + \left( \frac{5}{3} \Upsilon + \cL + \frac{1}{6} \cU' +\frac{1}{36}\cU^2  \right) \nu_- + \frac{1}{3} \cU \nu_-' + \nu_-'' + \ldots,  \nonumber \\ \\
\label{22}
&&\delta \bar{\varphi}_+ = \frac{3}{2} \varphi_+ \xi'\, + \varphi_+' \xi \, - \frac{1}{6}\cU \varphi_+ \xi + \frac{1}{6}\varphi_+ \eta + \left( \frac{5}{3} \Upsilon + \cL - \frac{1}{6} \cU' +\frac{1}{36}\cU^2  \right) \nu_- + \frac{1}{3} \cU \nu_-' + \nu_-''+\ldots,\nonumber \\ \\
\label{33}
&&\delta \cL = \frac{1}{2}\xi''' + 2\cL \xi' + \cL' \xi - \left(  \frac{1}{18}\cU \bar{\varphi}_+ + \frac{1}{6} \bar{\varphi}'_+ + \frac{5}{4\sqrt{6}}\bar{\Psi}_+ \right)\nu_- - \frac{1}{2}\bar{\varphi}_+ \nu'_- + \ldots \\
&&\delta \Upsilon = \Upsilon' \xi + 2\Upsilon \xi' + \sqrt{\frac{3}{32}}\left( \bar{\Psi}_+ \nu_- - \Psi_+ \bar{\nu}_- \right) + \ldots \\
\label{44}
&&\delta \cU = \eta' + \bar{\varphi}_+ \nu_{-} - \varphi_+ \bar{\nu}_{-} + \ldots 
\eea
where we refer the reader to Appendix \ref{AppB} for the ellipses which are not important for the following discussion. After some algebra, we found that the suitable field/parameter redefinitions are as follows (hatted variables are the new ones): 
\bea
&&\hat{\varphi}_+ =\frac{c}{3} \varphi_+,\,\,\,\hat{\varphi}_- =\frac{c}{3} \bar{\varphi}_+,\,\,\, \cr
&&\hat{U} = -\frac{c}{18}\, \cU, \cr
&&\hat{T} = \frac{c}{6} \left(  \cL + \frac{5}{3} \Upsilon + \frac{1}{36} \cU^2   \right) \cr
&&\hat{\eta}= - \frac{1}{6} \left( \eta + \frac{18}{c} \left( \cU\, \xi \right)'  \right) 
\eea
with $c$ a constant which, as we shall see shortly, has the meaning of the central charge. Upon the above redefinitions, we found that \eqref{11}-\eqref{44} can be written as (the ellipses refer to other terms unimportant for this particular computation)
\bea
\label{U0v}
\delta \hat{U} &=& \frac{c}{3} \hat{\eta}' + \hat{U}' \xi + \hat{U} \xi' + \cdots \\
\label{Tv}
\delta \hat{T} &=& \frac{c}{12} \xi''' + 2\hat{T}\xi' + \hat{T}' \xi + \cdots \\
\label{varphiv}
\delta \hat{\varphi}_+ &=& \frac{3}{2} \hat{\varphi}_+ \xi' + \hat{\varphi}'_+ \xi +
\hat{\varphi}_+\,\hat{\eta} + \frac{c}{3} \nu_-'' - 2 \hat{U} \nu_-' +  \left( 2\hat{T} - \hat{U}'   \right) \nu_- + \cdots\\
\label{varphimv}
\delta \hat{\varphi}_- &=& \frac{3}{2} \hat{\varphi}_- \xi' + \hat{\varphi}'_- \xi -
\hat{\varphi}_-\,\hat{\eta} + \frac{c}{3} \bar{\nu}_-'' - 2 \hat{U} \bar{\nu}_-' +  \left( 2\hat{T} + \hat{U}'   \right) \bar{\nu}_- \cdots
\eea
The relevant Noether current which we denote as $J_s$ reads
\be
J_{s} = \frac{1}{2\pi} \left( \xi \hat{T}+ \hat{\eta} \hat{U} + \nu_- \hat{\varphi}_- + \bar{\nu}_-\hat{\varphi}_+  \right)
\ee
and it generates the variations \eqref{U0v}-\eqref{varphimv}. Invoking Cauchy's residue theorem and \eqref{noether} yields the following OPEs:
\bea
\hat{T}(z) \hat{T}(0) &\sim& \frac{c}{2z^4} + \frac{2\hat{T}}{z^2} + \frac{\partial \hat{T}}{z} \\
\hat{T}(z) \hat{U}(0) &\sim& \frac{\hat{U}}{z^2} + \frac{\partial \hat{U}}{z} \\
\hat{T}(z) \hat{\varphi}_{\pm}(0) &\sim&\frac{3\hat{\varphi}_{\pm}}{2z^2} + \frac{\partial \hat{\varphi}_{\pm}}{z} \\
\hat{\varphi}_{\pm}(z) \hat{\varphi}_{\mp}(0) &\sim&  \frac{2c}{3z^3} + \frac{2\hat{T}}{z} \pm \frac{2\hat{U}}{z^2} \pm \frac{\partial \hat{U}}{z} \\
\hat{U}(z) \hat{\varphi}_{\pm}(0) &\sim& \pm \frac{\hat{\varphi}_{\pm}(0)}{z} \\
\hat{U}(z) \hat{U}(0) &\sim& \frac{c}{3z^2}.
\eea 
From the OPEs, we see that the constant $c$ can be identified as the central charge. One can expand the fields in terms of their Laurent modes 
\be
\hat{T} = \sum_n \frac{\hat{L}_n}{z^{n+2}},\,\,
\hat{\varphi}^{\pm} = \sum_r \frac{\hat{\varphi}^{\pm}_r}{z^{r+3/2}},\,\,\hat{U} = \sum_n \frac{\hat{U}_n}{z^{n+1}},\,\,
\ee
upon which the OPEs lead to the following commutation relations displayed below for completeness.
\bea
\label{q}
[\hat{L}_m, \hat{\varphi}^{\pm}_n ] &=& \left( \frac{1}{2}m - n \right) \hat{\varphi}^{\pm}_{m+n} \cr
[\hat{L}_m, \hat{U}_n ] &=& -n \hat{U}_{m+n}\cr
[\hat{U}_m, \hat{U}_n ] &=& \frac{1}{3}c m \delta_{m+n,0}\cr 
[\hat{U}_m, \hat{\varphi}^{\pm}_r ] &=& \pm \hat{\varphi}^{\pm}_{m+r}   \cr 
\{\hat{\varphi}^{\pm}_r, \hat{\varphi}^{\mp}_s \} &=& 2\hat{L}_{r+s} \pm (r-s) \hat{U}_{r+s} + \frac{c}{3} \left( r^2 - \frac{1}{4} \right) \delta_{r+s,0} 
\eea
We can immediately recognize \eqref{q} as the $\cN=2$ super-Virasoro algebra. Apart from being a good consistency check on our computations, the above exercise demonstrates that the Sugawara redefinition of the energy-momentum tensor should read as
\be
\label{energy}
\hat{T} = \frac{c}{6} \left(  \cL + \frac{5}{3} \Upsilon + \frac{1}{36} \cU^2   \right) = \frac{c}{36} \text{str} \left( a^2 \right)
\ee
where the gauge connection $a$ takes the form in \eqref{gaugec}.\footnote{The second equality in \eqref{energy} relies on inserting a factor of `$i$' in the generator $U_0$. As explained later, it turns out this is also required by a consistent reduction of this theory to $osp(N|2)$ theories.} Finally, we recall that the $\cN=2$ superconformal algebra enjoys the following spectral flow as an automorphism \cite{Schwimmer:1986mf}
\bea
\label{spectralflow}
\hat{U}_m &\rightarrow& \hat{U}_m + \frac{1}{3} \alpha c \delta_{m,0},\cr
\hat{\varphi}^{\pm}_r &\rightarrow& \hat{\varphi}^{\pm}_{r \pm \alpha},\cr
\hat{L}_m &\rightarrow& \hat{L}_m + \alpha \hat{U}_m + \frac{1}{6} \alpha^2 c \delta_{m,0}.
\eea
Later, we will see this automorphism manifest when we discuss some flat connections in the bulk Chern-Simons theory. A constant shift of the field conjugate to the generator $U_0$ induces a phase shift in the Killing spinor, and modifies the energy-momentum tensor following \eqref{energy}.
\section{Supersymmetry of $sl(3|2)$ Chern-Simons theory}\label{sec:groundstates}
\label{sec:four}
\subsection{A class of gauge transformations}
\label{sec:gaugetransform}

We now turn to the subject of deriving the suitable supersymmetry transformation laws for the $sl(3|2)$ Chern-Simons theory. Our consideration below lies in the chiral sector, but applies equally to the anti-chiral sector. Now, the gauge connection $\Gamma$ is $sl(3|2)$-valued and parametrized as
\be
\label{superc}
\Gamma = (e^a+\omega^a) J_a + \Upsilon^i K_i + \cU\,U_0 + \cW^m W_m + \psi^{(\frac{1}{2})}_r Q^{(\frac{1}{2})}_r + \bar{\psi}^{(\frac{1}{2})}_r \bar{Q}^{(\frac{1}{2})}_r + \psi^{(\frac{3}{2})}_r Q^{(\frac{3}{2})}_r + \bar{\psi}^{(\frac{3}{2})}_r \bar{Q}^{(\frac{3}{2})}_r 
\ee
with 
$$
K_2=A_0, \qquad K_1 = \frac{1}{2} \left( A_{1} - A_{-1}  \right), \qquad K_0 = \frac{1}{2} \left( A_{1} + A_{-1} \right)\,.
$$
Also, we should relate the gravitational vielbeins $e^a$ and spin connection $\omega^a=\frac{1}{2}\epsilon^{abc} \omega_{bc}$ to the relevant fields in the other copy of Chern-Simons by setting
\be
\label{ordgauge}
\tilde{\Gamma}= (-e^a + \omega^a)J_a + \ldots 
\ee 
Now, it is useful to begin by considering the invariance of the action under a gauge transformation that is valued in $Q^{(\frac{1}{2})}, \bar{Q}^{(\frac{1}{2})}$ (in the notation introduced in the previous section), i.e.
\be
\label{susyepsilonosp}
\epsilon_{susy} = \epsilon_r Q^{(\frac{1}{2})}_r + \bar{\epsilon}_s \bar{Q}^{(\frac{1}{2})}_s
\ee
Then, the invariance of the action under $\delta \Gamma$ which reads
\be
\label{supergauge}
\delta \Gamma = d\epsilon_{susy} + \left[ \Gamma, \epsilon_{susy} \right]
\ee
is guaranteed up to total derivative terms. From \eqref{supergauge}, we can then compute the supersymmetry transformation laws purely from the superalgebra. This method relies on the coincidence that we can write the gravity theory as a Chern-Simons theory of which bulk action is gauge invariant, modulo total derivatives which may not vanish at the boundary. Equation \eqref{supergauge} can then be used to derive the supersymmetry transformation laws, based on the $sl(2|1) \simeq osp(2|2)$ subalgebra, albeit a subtlety that involves the reality conditions of the fermionic fields. We note that when written in components, the kinetic terms read schematically as 
\bea
\cL_{kin.} &=& (e^a+\omega^a) \wedge d (e_a+\omega_a) + 2(e^a+\omega^a)\wedge d\Upsilon_a + \Upsilon^a \wedge d\Upsilon_a + \cU \wedge d\cU + \cW^m \wedge d\cW_m \cr
&& -  \bar{\psi}^{(\frac{1}{2})} \gamma_0 \wedge d\psi^{(\frac{1}{2})} -  \psi^{(\frac{3}{2})}\tilde{\gamma_0} \wedge d\bar{\psi}^{(\frac{3}{2})},
\eea
where the indices are contracted via the Killing metric of $sl(3|2)$, and 
$$
\tilde{\gamma}_0 = \begin{pmatrix}
0 & 0 & 0 & -1\\
0 & 0 & 1 & 0\\
0 & -1 & 0 & 0\\
1 & 0 & 0 & 0\\
\end{pmatrix}.
$$
Now, there is a crucial yet subtle point that affects how one derives the supersymmetry transformation laws. We note that if $\bar{\psi}$ and $\psi$ are independent, real Grassmann fields, then their kinetic terms are not real. As in the case of $osp(N|M)$ Chern-Simons theories, one inserts extra factors of `$i$' in the Lagrangian appropriately. 

In the following, we will keep track of these insertions by imposing suitable reality conditions on the fermionic fields $\psi$ as follows.
\be
\label{realitychange}
\bar{\psi}^{(\frac{1}{2})}_r = i   \psi^{(\frac{1}{2})\dagger}_r,\qquad
\bar{\psi}^{(\frac{3}{2})}_r = -i    \psi^{(\frac{3}{2})\dagger}_r.
\ee
We should note that the number of degrees of freedom involving $\psi$ remains the same, and that the sub-superalgebra generated by the generators $L_i, U_0, Q^{(\frac{1}{2})}_r, \bar{Q}^{(\frac{1}{2})}_r$ closes, and can be identified as $sl(2|1)\simeq osp(2|2)$. More precisely, we find that these algebras are isomorphic upon the identifications (please see Appendix \ref{AppendixA} for the generators of $osp(2|2)$ displayed below):
\bea
&&E \sim L_-,\,\,\, F \sim -L_+,\,\,\, H \sim 2L_0, \cr \cr
\label{sl32a}
&&J_{12} \sim i6U_0,\,\,\, Q^{(\frac{1}{2})}_{\pm \frac{1}{2}} \sim -\frac{i}{2\sqrt{3}} \left( R_1^{\pm} + i R_2^{\pm}  \right),\,\,\,\bar{Q}^{(\frac{1}{2})}_{\pm \frac{1}{2}} \sim \frac{1}{2\sqrt{3}} \left( R_1^{\pm} - i R_2^{\pm}  \right).
\eea
We note from \eqref{sl32a} that the matching between $sl(2|1)$ and $osp(2|2)$ involves changing the reality conditions of the generators $U_0$ and $Q^{(\frac{1}{2})}_{\pm \frac{1}{2}}$. These conditions ensure the reality of the superysymmetric Lagrangian, since they generate appropriate factors of `$i$' in the coefficients of the fermionic kinetic terms. Later, we shall use both \eqref{realitychange} and \eqref{supergauge} to derive the supersymmetry transformation laws, but first let us consider the cases where the fields are truncated to those in $osp(1|2)$ and $osp(2|2)$ Chern-Simons supergravity theories. The $sl(3|2)$ field content differs from that of the $osp(2|2)$ case by the additional $\cN=2$ multiplet $\{\Upsilon, \cW, \psi^{(3/2)}\}$. The supersymmetry transformation laws for these theories have been derived some time ago, and we want to re-derive them by using \eqref{supergauge} and \eqref{realitychange}, after setting the irrelevant fields in $sl(3|2)$ Chern-Simons theory to vanish. This should serve as a good consistency check of our approach. 


\subsection{On the supersymmetry of $osp(1|2)$ and $osp(2|2)$ Chern-Simons theories}
\label{sec:ospcheckone}

We begin with a comparison to $osp(1|2)$ Chern-Simons theory \cite{Achucarro:1987vz}. This theory can be recovered after a truncation of the $sl(3|2)$ theory, by setting 
\be
\label{trunosp1}
\cW = \Upsilon = \cU = \psi^{(\frac{3}{2})} = \bar{\psi}^{(\frac{3}{2})} = 0,
\ee
\be
\label{trunosp2}
\bar{\psi}^{(\frac{1}{2})} = -i\psi^{(\frac{1}{2})}.
\ee
To motivate \eqref{trunosp2}, we set the fields conjugate to the generator $R^{\pm}_2$ (see \eqref{sl32a} and Appendix \ref{AppendixA}) to be zero and the fields conjugate to $R^{\pm}_1$ to be real. After taking into account the overall trace normalization factor, we compute the action to be 
\footnote{We take the cosmological constant to be unity, and henceforth, rescaled all the fermonic fields, including the gauge parameters $\epsilon$ and $\bar{\epsilon}$ by a factor of $\sqrt{\frac{3}{2}}$.}
\bea
\label{actionosp1}
S &=& \frac{1}{8\pi G}\int d^3x \,\, e^a \wedge \left( d\omega_a + \frac{1}{2}\epsilon_{abc}\omega^b \wedge \omega^c  \right) + \frac{1}{6} \epsilon_{abc} \left( e^a \wedge e^b \wedge e^c   \right)  \cr
&&+ i \psi^{(\frac{1}{2})} \gamma_0 \wedge \left( d - \frac{1}{2} \left(e^a + \omega^a \right)\gamma_a  \right) \wedge \psi^{(\frac{1}{2})}
\eea
where the part consisting only of the vielbein and spin-connection one-forms is the usual Einstein-Hilbert action with a negative cosmological constant, and the fields $\psi^{(\frac{1}{2})}$ are real. This, including the factor of $i$, is the action for $osp(1|2)\oplus sl(2)$ Chern-Simons supergravity \cite{Achucarro:1987vz}. From \eqref{trunosp2} and \eqref{supergauge}, we obtain\footnote{Since we are supersymmetrizing only a copy of Chern-Simons theory, $\delta e = \frac{1}{2} \delta A$. The spin-connection one-forms are treated as auxiliary forms.} 
\bea
\label{osp1susy}
\delta e^a &=& \frac{1}{2} \epsilon \gamma_0 \gamma^a \psi^{(\frac{1}{2})} \\
\delta \psi^{(\frac{1}{2})} &=& \left( d - \frac{1}{2} \left( e^a + \omega^a   \right)\gamma_a \right) \epsilon
\eea
where we have also taken $\bar{\epsilon}=-i\epsilon$. We checked that \eqref{osp1susy} agrees with the supersymmetry transformation laws for $osp(1|2)$ Chern-Simons supergravity as stated in the literature. Similarly, we can consider the $osp(2|2) \oplus sl(2)$ Chern-Simons theory where the $u(1)$ gauge field $\cU$ plays a role. Instead of \eqref{trunosp1}, we set
\be
\label{trunosp3}
\cW = \Upsilon = \psi^{(\frac{3}{2})} = \bar{\psi}^{(\frac{3}{2})} = 0,
\ee
\be
\label{trunosp4}
\cU=i6\cB, \qquad  \bar{\psi}^{(\frac{1}{2})} = i  \psi^{(\frac{1}{2})\dagger}, \qquad 
\ee
which is motivated by constraining the fields conjugate to $R_{1,2}^{\pm}$ and $J_{12}$ to be real. This leads to the action
\bea
\label{actionosp1}
S &=& \frac{1}{8\pi G}\int d^3x\,\,  e^a \wedge \left( d\omega_a + \frac{1}{2}\epsilon_{abc}\omega^b \wedge \omega^c  \right) + \frac{1}{6} \epsilon_{abc} \left( e^a \wedge e^b \wedge e^c   \right)  \cr
&&- i  \psi^{(\frac{1}{2})\dagger}  \gamma_0 \wedge \left( d - \frac{1}{2} \left(e^a + \omega^a \right)\gamma_a  + i \cB \right) \wedge \psi^{(\frac{1}{2})} + 2\cB \wedge d\cB,
\eea
where the fermonic fields are now complex, and thus possess twice the degrees of freedom as in the $osp(1|2)$ case. We checked that this is indeed the action for $osp(2|2) \oplus sl(2)$ Chern-Simons supergravity \cite{Izquierdo:1994jz} as stated in the literature. From \eqref{trunosp4} and \eqref{supergauge}, we obtain the supersysymmetry transformation laws to be
\bea
\label{osp2susy}
\delta e^a &=& -\frac{i}{4} \left(  \epsilon^{\dagger} \gamma_0 \gamma^a \psi^{(\frac{1}{2})} +   \psi^{(\frac{1}{2})\dagger} \gamma_0 \gamma^a    \epsilon \right)\\
\delta \psi^{(\frac{1}{2})} &=& \left( d - \frac{1}{2} \left( e^a + \omega^a   \right)\gamma_a +i\cB\right) \epsilon \\
\delta \cB &=& \frac{1}{4} \left( \epsilon^{\dagger} \gamma_0 \psi^{(\frac{1}{2})} + \psi^{(\frac{1}{2})\dagger} \gamma_0 \epsilon   \right).
\eea
We checked that \eqref{osp2susy} is indeed the transformation laws for $osp(2|2)$ Chern-Simons supergravity as stated in the literature \cite{Izquierdo:1994jz}. We now proceed to derive the supersymmetry transformation laws for $sl(3|2)$ Chern-Simons supergravity theory, based on \eqref{realitychange} and \eqref{supergauge}. The results of this subsection vindicated the consistency of such an approach when the fields are truncated to yield $osp(1|2)$ and $osp(2|2)$ theories.

We should remark that the equations for $\psi^{(\frac{3}{2})}$ in \eqref{realitychange} cannot be derived by seeking consistent truncations to the $osp(N|2)$ cases since in the first place, they are absent in these theories. But, as we shall observe next, this relation can be easily motivated by the form of some of the equations that we obtain from \eqref{supergauge}. Finally, let us summarize our approach in the following. One starts with $sl(3|2;\mathbb{R})$ superalgebra in deriving the Chern-Simons action, but then, insert factors of `$i$' into appropriate parts of the Lagrangian by hand to ensure reality. Starting with \eqref{realitychange} can be viewed as a way to keep track of these changes. The results of this subsection showed that in each sector, when the multiplet $(\Upsilon, \cW, \psi^{(\frac{3}{2})})$ is truncated, this procedure reduces the $sl(3|2)$ theory to the $osp(2|2)$ Chern-Simons theory.

\subsection{$\cN=2$ supersymmetry and Killing spinors}
\label{sec:killingspinors}

Upon substituting \eqref{superc} and \eqref{realitychange} into \eqref{supergauge}, excluding the equations for $\psi^{(\frac{3}{2})}$ in \eqref{realitychange}, and inserting factors of `$i$' in the generators $U_0$ and $\cW^m$, we find the following supersymmetry transformation laws 

\bea
\label{one}
\delta \psi^{(\frac{1}{2})} &=& \left( d + i\cU - \frac{1}{2}\left( (e^a+\omega^a) + \frac{5}{3} \Upsilon^a  \right) \gamma_a \right) \epsilon \\
\label{two}
\delta \bar{\psi}^{(\frac{1}{2})} &=& \left( d - i\cU - \frac{1}{2}\left( (e^a+\omega^a) + \frac{5}{3} \Upsilon^a  \right) \gamma_a \right) \bar{\epsilon} \\
\label{three}
\delta \psi^{(3/2)} &=& \left( i\cW^m \eta_m + \Upsilon^a  \lambda_a \right) \epsilon \\
\label{four}
\delta \bar{\psi}^{(3/2)} &=& \left( i\cW^m \eta_m - \Upsilon^a  \lambda_a \right) \bar{\epsilon} \\
\label{five}
\delta \cU &=& \frac{1}{4} \left( \epsilon^{\dagger} \gamma_0 \psi^{(\frac{1}{2})} + \psi^{(\frac{1}{2})\dagger} \gamma_0 \epsilon   \right) \\
\label{six}
\delta \cW^m &=& \frac{i}{\sqrt{8}} \left(   \bar{\psi}^{(\frac{3}{2})} \alpha^m \epsilon +  \psi^{(\frac{3}{2})} \alpha^m \bar{\epsilon} \right) \\
\label{seven}
\delta e^a &=& -\frac{5}{6} \left( \delta \Upsilon^a  \right) -\frac{i}{4} \left(  \epsilon^{\dagger} \gamma_0 \gamma^a \psi^{(\frac{1}{2})} +   \psi^{(\frac{1}{2})\dagger} \gamma_0 \gamma^a    \epsilon \right)\\
\label{eight}
\delta \Upsilon^a &=&   \bar{\psi}^{(\frac{3}{2})} \beta^a \epsilon - \psi^{(\frac{3}{2})} \beta^a \bar{\epsilon}\,,
\eea
where the field $\cU$ is multiplied by a factor of 6, and the field $\psi^{(\frac{1}{2})}$ by a factor of $\sqrt{\frac{3}{2}}$ as in Section \ref{sec:ospcheckone} for convenience. The non-vanishing elements of the $4\times 2$ matrices $\beta, \eta, \lambda, \alpha$ are
$$
\alpha^0_{(\frac{1}{2},-\frac{1}{2})}=\alpha^0_{(-\frac{1}{2},\frac{1}{2})}=-\alpha^{-1}_{(\frac{1}{2},\frac{1}{2})}=-\alpha^{1}_{(-\frac{1}{2},-\frac{1}{2})}=1.\qquad 
\alpha^{-2}_{(\frac{3}{2},\frac{1}{2})}=\alpha^2_{(-\frac{3}{2},-\frac{1}{2})}=-\alpha^{-1}_{(\frac{3}{2},-\frac{1}{2})}=-\alpha^{1}_{(-\frac{3}{2},\frac{1}{2})}=\sqrt{\frac{1}{3}}.
$$
$$
\lambda^2_{(\frac{1}{2},\frac{1}{2})}=\lambda^2_{(-\frac{1}{2},-\frac{1}{2})}=-\frac{2\sqrt{2}}{3},\,\,   
-\lambda^{1}_{(\frac{3}{2},\frac{1}{2})}=\lambda^{1}_{(-\frac{3}{2},-\frac{1}{2})}=\sqrt{\frac{2}{3}},\,\,
-\lambda^{1}_{(\frac{1}{2},-\frac{1}{2})}=\lambda^1_{(-\frac{1}{2},\frac{1}{2})}=\frac{\sqrt{2}}{3}.
$$
$$
\lambda^{0}_{(\frac{3}{2},\frac{1}{2})}=\lambda^{0}_{(-\frac{3}{2},-\frac{1}{2})}=\sqrt{\frac{2}{3}},\,\,
\lambda^{0}_{(\frac{1}{2},-\frac{1}{2})}=\lambda^0_{(-\frac{1}{2},\frac{1}{2})}=\frac{\sqrt{2}}{3}.
$$
$$
\eta_m= \frac{1}{\sqrt{8}} \tilde{\gamma}_0\,\, \text{str}\left( W_m W_n  \right) \alpha^n,\qquad \beta_a=-i\frac{3}{8} \tilde{\gamma}_0 \lambda^a.
$$
Equation \eqref{four} suggests we should look for a reality condition relating $\bar{\psi}^{(\frac{3}{2})}$ and $\psi^{(\frac{3}{2})}$ since $\bar{\epsilon}=i\epsilon^{\dagger}$. A simple observation tells us that imposing
\be
\bar{\psi}^{(\frac{3}{2})}_r = -i \psi^{(\frac{3}{2})\dagger}_r  
\ee
throughout is consistent with \eqref{three} and \eqref{four}. This then completes the derivation of \eqref{realitychange}. We should recall again that this analytic continuation is accompanied by inserting `$i$' in the generators $U_0$ and $W^m$. Now, the Killing spinor equations can then be directly read off from \eqref{one}$-$\eqref{four} to be
\be
\label{spinor1}
\left( d + i \cU - \frac{1}{2}\left( (e^a+\omega^a) + \frac{5}{3} \Upsilon^a  \right) \gamma_a \right) \epsilon = 0,
\ee
\be
\label{spinor2}
\left( i\cW^m \eta_m + \Upsilon^a  \lambda_a \right) \epsilon = 0.
\ee
As we shall observe later, if we demand all fields to be real-valued, \eqref{spinor2} presents a rather stringent condition on the higher-spin fields that are allowed for a non-vanishing two-component complex spinor $\epsilon$. These results apply equally to the anti-chiral sector.



\subsection{Solving the Killing spinor equations}
\label{sec:solvk}
As a warm-up, we begin with an ansatz for a class of flat connections with the vielbein and spin connection as the only non-vanishing fields,
\bea
\label{basicconnection}
\Gamma&=&\left( e^{\rho} L_1 - \cL e^{-\rho} L_{-1} \right) dx^+ + L_0\, d\rho \\
\label{antibasicconnection}
\tilde{\Gamma}&=&-\left( e^{\rho} L_{-1} - \tilde{\cL} e^{-\rho} L_{1} \right) dx^- - L_0\, d\rho.
\eea
In ordinary gravity where the gauge group is $sl(2) \oplus sl(2)$, the parameters $\cL, \tilde{\cL}$ are related
to the ADM mass $M$ and angular momentum $J$ via the following equations
\be
\label{ADMquantities}
\cL= \frac{M-J}{2k},\qquad \tilde{\cL} = \frac{M+J}{2k}.
\ee
In particular, global $AdS_3$ corresponds to taking $\cL = \tilde{\cL} = -\frac{1}{4}$. Extremal black holes with positive $J$ correspond to taking $\cL=0, \tilde{\cL}>0$, while those with negative $J$ correspond to taking $\tilde{\cL} = 0, \cL>0$.

Consider the copy of Chern-Simons theory parametrized by flat connections $\Gamma$. The corresponding Killing spinor equations read 
\be
\label{k1}
\left( \partial_+ - e^{\rho} \gamma_+ + \cL e^{-\rho} \gamma_-    \right) \epsilon = 0, \qquad
\left( \partial_{\rho} - \frac{1}{2} \gamma_2  \right) \epsilon = 0, \qquad \partial_- \epsilon = 0.
\ee
where we have defined
\be
\gamma_{\pm} \equiv \frac{1}{2} \left( \gamma_0 \pm \gamma_1  \right).
\ee
If we let the spinor $\epsilon$ to be variable separable in $\rho, x^{\pm}$, then \eqref{k1} gives us
\be
\label{k1result}
\left( \begin{array}{c} e^{-\frac{\rho}{2}} \partial_+ K(x^+) \\ e^{\frac{\rho}{2}} K(x^+)
\end{array} \right),\qquad \partial^2_+ K= \cL K.
\ee
This implies the following classification: for $\cL > 0$, since $K$ is a sum of hyperbolic functions which are not periodic in $x^+$, there is no admissable Killing spinors. For $\cL =0$, $K\sim c_1 x^{+} + c_2$ where $c_{1,2}$ are constants. We have to set $c_1 = 0$ to preserve the periodicity, and hence the Killing spinor preserved is of the form 
\be
\label{k1result}
\epsilon_{\cL =0 } \sim \left( \begin{array} {c} 0 \\ e^{\frac{\rho}{2}}
\end{array} \right).
\ee
Finally, for negative $\cL$,  we have
\be
K= c_1 \sin \left(  \sqrt{|\cL|} x^+ \right)  + c_2 \cos \left( \sqrt{|\cL|} x^+ \right).
\ee
These are periodic for $|\cL | = N^2, N\in \mathbb{Z}^+$ and anti-periodic if $|\cL | = 
\left( \frac{2N+1}{2} \right)^2 $. In particular, if we restrict ourselves to $\cL \geq -\frac{1}{4}$, then the only supersymmetric case corresponds to global $AdS_3$ which will have two linearly independent $\epsilon$'s. Identical results hold for $\bar{\epsilon}$, and thus the number of real Killing spinors preserved in each case is to be doubled. 

We can also supersymmetrize the other copy of Chern-Simons theory, and obtain similar results depending on the sign of $\bar{\cL}$. We found that substituting \eqref{antibasicconnection} into \eqref{spinor1} amounts to switching 
\be
\label{switch}
x^{\pm} \rightarrow x^{\mp},\qquad  \gamma_2 \rightarrow - \gamma_2,\,\gamma_{\pm} \rightarrow - \gamma_{\mp}.
\ee
Instead of \eqref{k1result}, the Kiling spinors are
\be
\label{k2result}
\left( \begin{array}{c} e^{\frac{\rho}{2}} \tilde{K}(x^-) \\ e^{-\frac{\rho}{2}} \partial_- \tilde{K}(x^-)
\end{array} \right),\qquad \partial^2_- \tilde{K}= \tilde{\cL}\, \tilde{K}.
\ee
This implies that in the $\cN=(2,2)$ theory based on $sl(3|2) \oplus sl(3|2)$, global $AdS_3$ ($\cL=\tilde{\cL}=-\frac{1}{4}$) preserves 8 real supercharges, the massless BTZ preserves 4, and extremal black holes with non-zero angular momentum preserve 2. The generic BTZ with $J\neq M$ will break all supersymmetries. These results agree with those belonging to the case of $osp(2|2)\oplus osp(2|2)$ Chern-Simons supergravity theories \cite{Coussaert:1993jp}.

To be more general, given any $x^{\pm}$-component of the gauge connection $\Gamma$, it is straightforward to solve for the form of any admissable Killing spinors as:
\bea
\label{generalspinor}
\epsilon &=& e^{\left(-i\cU + m_0\right) x  } \bigg[ \alpha e^{A_+ x}   \left( \begin{array}{c} A_+ \\ m_+ \\
\end{array} \right) + \beta e^{A_- x}   \left( \begin{array}{c} A_- \\ m_+ \\
\end{array} \right)
   \bigg] \\
A_{\pm} &=& -m_+m_0 \pm \sqrt{m_+^2m_0^2 + m_+m_-},\qquad m_i \equiv \left(e^i + \omega^i + \frac{5}{3}\Upsilon^i \right),\,\,x\equiv x^{\pm}\,,
\eea
with $\alpha, \beta$ being arbitrary complex constants. The remaining spinor equation 
\eqref{spinor2} constrains the Killing spinors to lie within the null-space of the following matrix of one-forms $\cW^m$ and $\Upsilon^m$:
\be
\label{projectable}
\left( \begin{array}{cc} -i\cW^{-1} + 2\Upsilon^{-1} & -4i\cW^{-2} \\ i\cW^0 - \Upsilon^0 & \frac{3}{2} i\cW^{-1} + \Upsilon^{-1} \\ -\frac{3}{2} i\cW^{1} + \Upsilon^{1} & -i\cW^0 - \Upsilon^0 \\ 4i\cW^2 & i\cW^1 + 2\Upsilon^1 \\
\end{array} \right).
\ee
This turns out to be highly restrictive on the fields $\cW$ and $\Upsilon$ in the gauge connection, apart from the periodicity conditions that one should further impose on the spinors in \eqref{generalspinor}. 


\subsection{On the $u(1)$ gauge field}
\label{sec:u1g}
As mentioned earlier, it is well-known that $\cN =2$ superconformal algebra admits an automorphism, with the spectral flow generated by the zero mode of the $u(1)$ field. We expect to find this automorphism symmetry in the bulk Chern-Simons theory. Let us first consider the chiral sector with a generic gauge connection $\Gamma$ that contains only even-graded generators. It is easy to observe that a constant shift of the $u(1)$ field component in the $x^+$ direction 
$$
\cU \rightarrow \cU + \delta\cU
$$
can be realized by performing a large gauge transformation 
\be
\label{largeg}
\Gamma \rightarrow \Gamma + e^{-i\delta \cU\,U_0 x^+}de^{i\delta \cU\,U_0 x^+},
\ee
since $U_0$ commutes with all the even-graded generators. Also, as observed earlier, any Killing spinor attains a phase factor 
\be
\label{phasef}
\epsilon \rightarrow e^{-i\delta \cU x^+}\epsilon.
\ee
Similarly, for the anti-chiral sector, we have
\be
\label{largeg}
\tilde{\Gamma} \rightarrow \tilde{\Gamma} + e^{i\delta \tilde{\cU}\,U_0 x^-}de^{-i\delta \tilde{\cU}\,U_0 x^-},\qquad
\tilde{\epsilon} \rightarrow e^{i\delta \tilde{\cU} x^-}\tilde{\epsilon}.
\ee
From \eqref{energy}, we can compute the changes to the energy-momentum tensor
\be
\label{changeE}
T_{++} \rightarrow T_{++} + 2\delta \cU + \delta \cU^2, \qquad \tilde{T}_{--} \rightarrow \tilde{T}_{--} + 2\delta \tilde{\cU} + \delta \tilde{\cU}^2.
\ee
Starting from any generic solution with zero $u(1)$ charge, one can generate a family of solutions. For supersymmetric ones which preserve Killing spinors, the $u(1)$ charge is thus quantized as, in our choice of normalization, 
\be
\label{Rchargeq}
\cU \in \mathbb{Z}/2,\qquad \tilde{\cU} \in \mathbb{Z}/2.
\ee  
It turns out that this quantization is also consistent with the requirement of a smooth holonomy. As an explicit example, consider the ansatz \eqref{basicconnection} and \eqref{antibasicconnection}, and recall from earlier discussion that for supersymmetry to be preserved, we require $\cL \leq 0$. This implies the existence of a family of supersymmetric solutions satisfying
\be
\label{Ufieldsusysol}
\sqrt{|\cL|} + \cU  \in \mathbb{Z}/2, \qquad \sqrt{|\tilde{\cL}|} + \tilde{\cU}  \in \mathbb{Z}/2.
\ee
It is straightforward to check that imposing a trivial holonomy (more about holonomy conditions in the next section) along the 
$\phi$-direction for this class of solutions yields both \eqref{Ufieldsusysol} and \eqref{Rchargeq}. Also, generalizing \eqref{ADMquantities}, the physical charges of mass ($M$) and angular momentum ($J$) read
\bea
\label{MandJ}
M &=& k \left( \cL + \tilde{\cL} +  \cU^2 + \tilde{\cU}^2 \right) \\
J &=& k \left( -\cL + \tilde{\cL} -\cU^2 + \tilde{\cU}^2 \right).
\eea

In supergravity theories that arise from type IIB string theory compactified on some internal space, the conical defect spacetimes in ordinary 3d gravity, with masses interpolating between $AdS_3$ and the massless BTZ (i.e. in our notation, $-\frac{1}{4} < \cL < 0$), can be embedded as solutions and made supersymmetric by turning on the $U(1)$ charge \cite{David, Mald, Boer}. Here, they are also supersymmetric solutions in the higher-spin supergravity theories, but we note that the trivial holonomy condition will be lost. The solutions in \eqref{Ufieldsusysol} correspond to conical surpluses instead, with $\cL < -\frac{1}{4}$. We will discuss more about conical defect spacetimes later in Section \ref{sec:onhol}.

\subsection{Holonomy conditions and higher-spin black holes}
\label{sec:holonomys}
In the $sl(N)$ Chern-Simons theory, higher-spin black holes and conical defects are defined via holonomy conditions. For black holes, the holonomy along the Euclidean time direction is trivial whereas for conical defects, the holonomy along the angular direction $\phi$ is trivial. In the $sl(3|2) \oplus sl(3|2)$ theory, consider first the case where only the fields $\cL, \tilde{\cL}$ survive. As shown in \ref{sec:killingspinors}, global $AdS_3$ and extremal BTZ are shown to be supersymmetric solutions in the theory. As reviewed earlier, their corresponding gauge connection $\Gamma$ can be parametrized as $\left(L_1 - \cL\,L_{-1}\right)$ of which eigenvalues of $\oint \Gamma_{\phi} d\phi$ read as $2\pi (0, \sqrt{\cL}, -\sqrt{\cL}, 2 \sqrt{\cL}, -2\sqrt{\cL})$, with $\cL=-1/4$ for global $AdS_3$.

If we wick-rotate the time-like direction $t \rightarrow -i\tau$, and let the Euclidean time period $(\Delta \tau)$ be such that the Euclidean manifold is smooth, then for the BTZ black hole, we have $\Delta \tau = \frac{\pi}{\sqrt{\cL}}$. For thermal $AdS_3$ or the Euclidean BTZ, the holonomy along $\tau$-direction reads
\be
\label{holisl3}
e^{i\oint \Gamma d\tau} \sim  \left[ \begin{array}{c|c} \mathbf{1}_{3\times 3} & 0 \\ \hline
0 & -\mathbf{1}_{2\times 2} \end{array}\right].
\ee
We should note that although \eqref{holisl3} is not the center of $sl(3|2)$, it is a central element among all group elements derived from exponentiating even-graded generators. To see this, we note that \eqref{holisl3} is a linear combination of the identity and $U_0$. On the other hand the holonomy, $e^{\oint \Gamma d\phi}$ possesses the eigenvalues $(1, e^{2\pi  \sqrt{\cL}}, e^{-2\pi  \sqrt{\cL}}, e^{ 4\pi \sqrt{\cL}}, e^{-4\pi  \sqrt{\cL}}   )$. In the limit $\cL \rightarrow 0^+$, the holonomy along the $\phi$-direction becomes precisely the identity (the Euclidean-time direction becomes non-compact in this limit). This is the holonomy condition for the massless BTZ. An important result of \cite{Kraus1} is that for $sl(N)$ higher-spin black holes constructed in this manner, the holonomy conditions coincide with the integrability conditions (which we shall review in a moment) that point towards the existence of a partition function containing the boundary source terms. 

To find appropriate generalizations of the higher-spin $sl(3)$ black holes in the $sl(3|2)$ theory, it seems natural to begin by considering an ansatz where $\Gamma_+$ is in the highest-weight gauge\footnote{We set the abelian field $\cU =0$ for simplicity in this Section.}. Paralleling the approach in \cite{Kraus1}, we begin with
\be
\label{connectiong}
\Gamma = b^{-1} a(x^+) b + b^{-1} db,\qquad \bar{\Gamma} = b\bar{a}(x^-) b^{-1} + bdb^{-1}
\ee
where $b=e^{\rho L_0}$, and
\bea
\label{gansatz1}
a &=& \left( L_1 - \cL L_{-1} + \cW W_{-2} +  \Upsilon A_{-1} \right) dx^+  + \left( \sum_{i=-2}^{2} \chi_i W_i + \sum_{i=-1}^{1} \xi_i L_i + \sum_{i=-1}^{1} \cA_i A_i \right) dx^-,\cr
\bar{a} &=& -\left( L_{-1} -  \bar{\cL} L_1 +  \bar{\cW} W_2 +  \bar{\Upsilon} A_1 \right) dx^- - \left( \sum_{i=-2}^{2} \bar{\chi}_i W_{-i} + \sum_{i=-1}^{1} \bar{\xi}_i L_{-i} + \sum_{i=-1}^{1} \bar{\cA}_i A_{-i} \right) dx^-.\nonumber \\
\eea
In the above, the unspecified functions are functions of the boundary co-ordinates ($x^{\pm}$). To furnish a prescription for bulk computations in the presence of source terms, the underlying principle is that the sources are associated with generalized boundary conditions for the various bulk higher-spin fields. Similar to the $sl(N)$ story, we wish to propose that functions $(\xi_1, \chi_2, \cA_1)$ are proportional to the sources in the putative CFT, and below, we will support this claim at the level of perturbation theory in the sources. This is done simply by comparing bulk field equations to the CFT's Ward identities, which will be performed explicitly below for the chiral sector. 

Now, the bulk field equations are just the conditions for a flat connection. For the ansatz \eqref{connectiong}, it is straightforward to solve for the various functions in terms of products of derivatives of the following set $(\mathcal{L}, \Upsilon, \mathcal{W}, \chi \equiv \chi_2, \xi \equiv \xi_1, \cA \equiv \cA_1 )$, in particular, with the fields $(\mathcal{L}, \Upsilon, \mathcal{W} )$ obeying the following differential equations.
\bea
\label{diffBulk}
\partial_{-}{\cL} &=& \frac{1}{2} \partial_+^{3} \xi + 2\cL \partial_+ \xi + \xi \partial_{+} \cL + \cA \partial_{+} \Upsilon + 2\Upsilon \partial_{+} \cA - 6\cW \partial_{+} \chi - 4 \chi \partial_{+} \cW,\cr
\partial_{-}{\Upsilon} &=& \frac{1}{2} \partial_+^{3} \cA + 2\cL \partial_+ \cA + \cA \partial_{+} \cL + \xi \partial_{+} \Upsilon  + 2\Upsilon \partial_{+} \xi - 6\cW \partial_{+} \chi - 4 \chi \partial_{+} \cW,\\
\partial_{-} \cW &=& 3\cW (\partial_+ \xi + \partial_+ \cA) + (\xi + \cA) \partial_+ \cW +
\frac{8}{3} \left( (\cL + \Upsilon)^2 \partial_+ \chi  + \chi (\cL + \Upsilon)\partial_+ (\cL + \Upsilon)     \right) \cr
&&+\frac{1}{24} \partial_+^5 \chi + \frac{5}{6} (\cL+ \Upsilon) \partial_+^3 \chi + \frac{5}{4} \chi \partial_+(\cL + \Upsilon) + \frac{3}{4} \partial_+ \chi (\partial_+^2 \cL + \partial_+^2 \Upsilon) + \frac{1}{6}\chi (\partial_+^3 \cL + \partial_+^3 \Upsilon). \nonumber \\
\eea
On the other hand, from the perspective of the boundary CFT, upon the insertion of the source terms in the Lagrangian, in the form
\be
\label{source1}
\int d^2 x \left( \chi(x) \cW(x) + \cA(x) \Upsilon(x) + \xi(x) \cL(x) + \bar{\chi}(x) \bar{\cW}(x) + \bar{\cA}(x) \bar{\Upsilon}(x) + \bar{\xi}(x) \bar{\cL}(x)  \right)
\ee
the various expectation values of $(\cL, \Upsilon, \cW)$ will pick up the source terms due to the singular terms in the OPEs, as explained in \cite{Kraus1} and reviewed in Section \ref{sec:Wsymmetry}. We have assumed that the source terms are precisely the set of $(\xi, \cA, \chi)$ in the bulk ansatz. To validate this assumption, we need to ensure that the OPEs among the fields belong to that of the $\mathcal{N}=2\,\, \mathcal{W}_3$ algebra. To conveniently compare with the bulk equations \eqref{diffBulk}, we first switch to Euclidean co-ordinates $(z=it + \phi, \bar{z}=-it+\phi)$, and perform the re-scaling for the fields in the bulk ansatz
\be
\label{rescale}
\cL \rightarrow \alpha \cL, \Upsilon \rightarrow \beta \Upsilon, \cW \rightarrow \gamma \cW.
\ee
After some algebra, we find that \eqref{diffBulk} lead to the OPEs 
\bea
\label{OPEs}
\cL(z) \cL(0) &\sim& \frac{3}{\alpha z^4} + \frac{2\cL}{z^2} + \frac{\partial \cL}{z},\qquad
\Upsilon(z) \Upsilon(0) \sim \frac{3}{\beta z^4} + \left(  \frac{\alpha}{\beta} \right)   \left( \frac{2\cL}{z^2} + \frac{\partial \cL}{z} \right),\cr
\cL(z) \cW(0) &\sim& \frac{3\cW}{z^2} + \frac{\partial \cW}{z}, \qquad
\Upsilon(z) \cW(0) \sim \frac{3\cW}{z^2} + \frac{\partial \cW}{z}, \qquad
W(z) \Upsilon(0) \sim \left( -\frac{\gamma}{\beta} \right) \left( \frac{6\cW}{z^2} + \frac{4\cW'}{z}  \right),\cr
W(z) \cL(0) &\sim& \left( -\frac{\gamma}{\alpha} \right) \left( \frac{6\cW}{z^2} + \frac{4\cW'}{z}  \right)\qquad
\cL(z) \Upsilon(0) \sim \frac{2\Upsilon}{z^2} + \frac{\partial \Upsilon}{z}, \qquad
\Upsilon(z) \cL(0) \sim \left( \frac{\beta}{\alpha}  \right) \left(\frac{2\Upsilon}{z^2} + \frac{\partial \Upsilon}{z} \right),\cr \cr
\cW(z) \cW(0) &\sim& \frac{5}{\gamma z^6} + \frac{1}{6\gamma} \left( \frac{\partial^3(\alpha\cL + \beta\Upsilon)}{z} + \frac{9\partial^2(\alpha\cL + \beta\Upsilon)}{2z^2} + \frac{15\partial (\alpha\cL + \beta\Upsilon)}{z^3} + \frac{30 (\alpha\cL + \beta\Upsilon)}{z^4}       \right) \cr 
&&+\frac{4}{3\gamma} \left( \frac{2(\alpha\cL + \beta\Upsilon)^2}{z^2} + \frac{\partial (\alpha\cL + \beta\Upsilon)^2}{z} \right),
\eea
where in particular, we note the non-linear terms arising in the $\cW(z) \cW(0)$ OPE. Indeed, we find that we can choose appropriate scaling parameters $(\alpha,\beta,\gamma)$ such that \eqref{OPEs} are actually those of the $\mathcal{N} =2\,\,\cW_3$ algebra restricted to the bosonic fields (excluding the $U(1)$ field omitted for simplicity of discussion). For a precise comparison, let us adopt the basis of the super $\cW_3$ algebra as constructed in reference \cite{Lu}. We find that upon choosing 
\be
\label{rescaling1}
\alpha = \beta = -2\gamma = \frac{3}{c}
\ee
and re-defining 
\be
T=\frac{1}{2} (\cL + \Upsilon), \qquad \bar{T}=\frac{1}{2} (\cL - \Upsilon), \qquad \tilde{W} = 2W,
\ee
the fields $(T,\bar{T},\tilde{W})$ generate the relevant OPEs in eqn. 19 of reference \cite{Lu}. In particular, we have established that $\chi$ acts as the source for the spin-3 field $\cW$. 

We now turn to the subject of constructing higher-spin black hole solutions, and investigating if the integrability of the higher-spin charges are related to the holonomy condition \eqref{holisl3}. Consider the same ansatz \eqref{gansatz1} but now with all the functions set to be constants. The connection now reads as\footnote{The factor of $-1/2$ was inserted in accordance with \eqref{rescaling1}.} 
\bea
\label{gansatz2}
a_+ &=&  L_1 - \cL L_{-1} - \frac{1}{2}\cW W_{-2} + \Upsilon A_{-1},\cr
a_- &=&  \bigg[\left(2\cW \chi + \Upsilon \cA \right)L_{-1} + \left( (\cL - \Upsilon )^2 \chi - \frac{1}{2}\cA \cW    \right) W_{-2} + \chi W_2 + 2 \chi \left( -\cL + \Upsilon  \right) W_0  \cr
&&- \left(-2 \cW \chi + \cL \cA    \right) A_{-1} + \cA A_1   \bigg] 
\eea
where, for simplicity of discussion, we have set $\xi$ - the source for $\cL$ - to be zero. Apart from $\cL$, the parameter space of the ansatz consists of the fields $\cW, \Upsilon$ and their conjugate potentials $\chi, \cA$. 

What are the `integrability conditions'? Now, in the context of the solution as described in \eqref{gansatz2}, the chemical potentials for the fields $\cW$ and $\Upsilon$ read as 
$$\mu_{\cW} \equiv -\tau \chi,\qquad \mu_{\Upsilon} \equiv -\tau \cA$$
respectively, up to some normalization. The integrability conditions refer to the relations
\be
\label{integrability}
\frac{\partial \cL}{\partial \mu_{\cW}} = \frac{\partial \cW}{\partial \tau},\qquad \frac{\partial \cL}{\partial \mu_{\Upsilon}} = \frac{\partial \Upsilon}{\partial \tau}, \qquad \frac{\partial \cW}{\partial \mu_{\Upsilon}} = \frac{\partial \Upsilon}{\partial \mu_{\cW}},
\ee
which is motivated by the existence of a partition function at the boundary that reads
\be
\label{parti}
Z \left(\tau, \mu_{\cW}, \mu_{\Upsilon} \right) = \text{Tr} \left(  e^{ k \left(\tau \cL + \mu_{\cW} \cW + \mu_{\Upsilon} \Upsilon - \bar{\tau} \bar{\cL} + \bar{\mu_{\cW}} \bar{\cW} + \bar{\mu_{\Upsilon}} \bar{\Upsilon} \right)}    \right)
\ee
where, for a static BTZ limit, the chiral and anti-chiral quantities are related as
\be
\tau = -\bar{\tau}, \mu_{\cW} = -\bar{\mu_{\cW}}, \mu_{\Upsilon} = -\bar{\mu_{\Upsilon}}, \cL = \bar{\cL}, \cW=-\bar{\cW}, \Upsilon = -\bar{\Upsilon}.
\ee
Let us now demand the holonomy to be that of \eqref{holisl3} and study its compatibility with \eqref{integrability}. For the solution \eqref{gansatz2}, after some algebra, we find that the five eigenvalues can be reduced nicely to the three roots of one cubic equation and the remaining two taking a rather nice form
\be
\label{twor}
\textrm{Eigenvalues} = \{\pm (\cA - 1) \sqrt{\cL + \Upsilon},\,\,\, \textrm{Roots of}\,\,x^3 -Bx + C= 0\}.
\ee
where $B, C$ are polynomial functions in $\{\cL,\cW,\Upsilon, \chi, \cA \}$, which read as
\be
\label{Beqn}
B\equiv -\frac{4}{3}\left( 16 \cL^2 \chi^2 - 3(1+\cA)^2 \Upsilon + 16\chi^2 \Upsilon^2 + \cL(3+6\cA +3\cA^2 - 32\chi^2 \Upsilon) + 18(1+2\cA)\chi \cW \right)
\ee
\bea
\label{Ceqn}
C&\equiv& 4 \bigg[ 256 \cL^3 \chi^3 - 144(1+\cA)^2 \chi \Upsilon^2 - 48\cL^2 \chi (3+6\cA + 3\cA^2 + 16\chi^2 \Upsilon) - 27(1+\cA)^2 (-1+2\cA) \cW \cr
&&+432 (3+2\cA)\chi^2 \Upsilon \cW + 48\cL \chi(6(1+\cA)^2 \Upsilon + 16\chi^2\Upsilon^2 - 9(3+2\cA)\chi \cW) - 128\chi^3 (2\Upsilon^3 + \cW^2) \bigg] \nonumber \\
\eea
We now impose the BTZ holonomy condition by solving for the fields $(\cL, \cW, \Upsilon)$ in terms of inverse temperature $\tau$, and potentials $\cA, \chi$. The `$\mathbf{1}_{2\times 2}$' factor in \eqref{holisl3} gives us conveniently, from \eqref{twor},
\be
\label{upsilone}
\Upsilon = \frac{\pi^2}{\tau^2 (1-\cA)^2} - \cL(\tau, \cA, \chi).
\ee
For the `$\mathbf{1}_{3\times 3}$' factor in \eqref{holisl3}, the zero eigenvalue demands setting $C=0$, which in turn gives us a quadratic equation for $\cW$. The other two eigenvalues of $(\pm 2\pi i)$ are then the roots of the cubic equation in \eqref{twor}, i.e.
\be
\label{Btau}
B=-\frac{4\pi^2}{\tau^2}.
\ee
Substituting the expressions for $\Upsilon$ and $\cW$ into $B$ in \eqref{Btau}, we can re-arrange the various terms to obtain a quartic equation for $\cL$. Although an analytic, closed form for $\cL$ is not manageable, we can perform a perturbative analysis to any desired order rather easily, and obtain $\cL$ (and thus also $\Upsilon$ and $\cW$) as a Taylor series in $(\cA, \chi)$. Below, we present the expressions for the various fields and the free energy up to some order.
\bea
\label{Fieldexpression}
\cL &=& \frac{\pi^2}{\tau^2} + \frac{40\pi^4\chi^2}{3\tau^4} + \frac{3\pi^2 \cA^2}{\tau^2} - \frac{80\pi^4\cA\chi^2}{\tau^4} + \frac{1280\pi^6 \chi^4}{3\tau^6} + \frac{280\pi^4 \chi^2 \cA^2}{\tau^4} + \frac{5\pi^2 \cA^4}{\tau^2} + \mathcal{O}(5)   ,\cr \cr
\cW &=& \frac{16\pi^4 \chi}{3\tau^4} - \frac{80\pi^4\chi \cA}{3\tau^4} + \frac{80\pi^4 \chi \cA^2}{\tau^4} + \frac{5120\pi^6\chi^3}{27\tau^6} -\frac{5120 \pi^6 \chi^3 \cA}{3\tau^6} - \frac{560 \pi^4 \chi \cA^3}{3\tau^4} +       \mathcal{O}(5)\cr \cr
\Upsilon &=& \frac{2\pi^2 \cA}{\tau^2} - \frac{40\pi^4 \chi^2}{3\tau^4} + \frac{4\pi^2 \cA^3}{\tau^2} + \frac{80 \pi^4 \chi^2 \cA}{\tau^4} - \frac{1280\pi^6 \chi^4}{3\tau^6} - \frac{280\pi^4 \chi^2 \cA^2}{\tau^4} +                      \mathcal{O}(5). \nonumber \\
\eea
It can then be checked that the integrability conditions of \eqref{integrability} are indeed satisfied, and that free energy in \eqref{parti} is integrable. Indeed, the free energy can be computed from \eqref{parti} and \eqref{Fieldexpression} after taking into account the contributions from the anti-chiral sector, and we obtain
\be
\label{free}
\text{ln}\, Z= -2 k \left( \frac{\pi^2}{\tau} + \frac{8\pi^4 \mu^2_{\cW} }{3\tau^5} + \frac{\pi^2 \mu^2_{\Upsilon}}{\tau^3} + \frac{40\pi^4 \mu^2_{\cW} \mu_{\Upsilon}}{3\tau^6} + \frac{40\pi^4 \mu^2_{\cW} \mu^2_{\Upsilon}  }{\tau^7} + \frac{\pi^2 \mu^4_{\Upsilon}}{\tau^5} \right) + \mathcal{O}(5)
\ee
If further, as first proposed in \cite{Kraus1}, the entropy is defined via a Legendre transform of the free energy, then the first law of black hole thermodynamics is naturally satisfied. Our black hole solutions thus provide a concrete evidence that defining higher-spin black holes via the BTZ holonomy condition, as originally proposed in the $sl(N)$ case \cite{Kraus1, Kraus2}, generalizes consistently to the $sl(N|N-1)$ theories. On the other hand, as briefly pointed out earlier, there exists a subtle difference with the $sl(N)$ theories, since \eqref{holisl3} is not the center of the group. One may ask if imposing the supermatrix identity as the holonomy condition can also be compatible with the integrability conditions, for example, by taking $\cL = - \Upsilon$ in \eqref{twor}, and whether there exists a gauge in which we can find a smooth horizon. We leave these questions for future investigations.

Finally, we point out that these solutions do not preserve supersymmetries due to the higher-spin fields they carry. In the next section, we will study a class of solutions that do preserve some amount of supersymmetry. It turns out that this requires us to strip off all the higher-spin fields except for the one conjugate to the lowest weight generator.

\subsection{Supersymmetric solutions with higher-spin fields}
\label{sec:susysolhigher}
If one demands the non-vanishing of any of the fields $\cW=\Upsilon=\chi=\cA=0$, then for the ansatz \eqref{gansatz2},  \eqref{projectable} implies $\Upsilon^{-1} \equiv\Upsilon= 0$, which forces either $\cW^{-2}\equiv \cW=0$ or the spinor component $\epsilon_2=0$, the latter condition fixing $\cL=0$. Solving the other two spinor equations and adding the $U(1)$ charge yields the following supersymmetric class of connections\footnote{We note that this class of solutions is the RG flow solution of \cite{Kraus2}, carrying some $U(1)$ charge. We thank an anonymous referee for his/her kind reminder.} 
\be
\label{Amm}
\Gamma = \left( e^{\rho} L_1 + 6\cU\, U_0 \right) dx^+ +  \mu e^{2\rho} W_2   dx^- + L_0\, d\rho\,,
\ee
\be
\label{barA}
\tilde{\Gamma} = \left( -e^{\rho} L_{-1} - 6\tilde{\cU} U_0 \right) dx^- - \tilde{\mu} e^{2\rho} W_{-2} dx^+ - L_0 d\rho.
\ee
Just as in the case of a vanishing $\mu$, demanding the holonomy $e^{\oint \Gamma d\phi}$ to be trivial turns out to be equivalent to \eqref{Rchargeq}. There is no additional constraint arising from the parameter $\mu$, but as in the non-supersymmetric theory, its non-vanishing implies that the metric grows as $e^{4\rho}$ and is thus asymptotic to $AdS_3$ with half its original $(\mu = 0)$ radius. Invoking the relation
\be
\label{metric} 
g_{\mu \nu} = \frac{1}{str(L_0^2)} \text{str} ( e_{\mu} \cdot e_{\nu} ), \qquad e = \frac{1}{2} \left( \Gamma - \tilde{\Gamma} \right),
\ee
we can compute the line element to be
\be
\label{metricbh}
ds^2 = d\rho^2 - \left( e^{2\rho} + \frac{16}{3} \mu \tilde{\mu} e^{4\rho} +2\cU \tilde{\cU} \right) dx^+\,dx^- +
\left( \cU^2 dx_+^2 + \tilde{\cU}^2 dx_{-}^2 \right).
\ee
To ensure the correct signature at infinity, we impose the parameter constraint
$$\mu \tilde{\mu} \geq 0.$$ 
Further, one can compute the spin-3 field $Q \sim \text{str} \left(e\cdot e \cdot e\right)$ (in the absence of $\cU$ and $\tilde{\cU}$) to read
\be
\label{spin3charge}
Q \sim e^{4\rho}\, \left( \tilde{\mu} dx_+^3 + \mu dx_-^3 \right).
\ee
As explained earlier, this class of solutions preserves two real supercharges in each chiral sector. We should also remark that \eqref{Amm} and \eqref{barA} can be further generalized to include a term $\sim W_2$ in $\Gamma_+$ (and correspondingly another term $\sim W_{-2}$ in $\bar{\Gamma}_-$), but including these terms destroy the asymptotically $AdS_3$ condition.

\section{Extension of results to $sl(N|N-1)$ for a general finite $N$.}
\label{sec:five}
\subsection{On the Killing spinor equations}
\label{sec:onKilling}
Recall from equation \eqref{groupstructure} how the general $sl(N|N-1)$ element can be decomposed as a direct sum of $sl(2)$ multiplets. In Appendix \ref{AppB}, we present explicit expressions for the generators and commutator relations for $sl(N|N-1)$. Below, we shall briefly present some important points following \cite{Fradkin:1990qk}. The gravitational $sl(2)$ is generated by $L^1_m$, of which the other generators are irreducible representations of. They can be expressed as 
\be
L_0 = \sqrt{\frac{N(N+1)}{12}} \left( \sqrt{N+2} T^1_0 + \sqrt{N-1} U^1_0   \right),\,\,\, L_{\pm 1} = \sqrt{\frac{N(N+1)}{6}} \left( \sqrt{N+2} T^1_{\pm 1} + \sqrt{N-1} U^1_{\pm 1}   \right),
\ee
while $V^1_m$ generates $h^{(1)}$ (another spin-1 multiplet of $sl(2)$). The spin-0 sector (i.e. $h^{(0)}$) is generated by 
\be
U = \sqrt{N} T^0_0 + \sqrt{N+1}U^0_0
\ee
which commutes with all even-graded generators. Together with $L_0, L_{\pm 1}$, they generate the even part of the $sl(2|1)$ sub-superalgebra, whereas $Q^{(\frac{1}{2})}_r$ and $\bar{Q}^{(\frac{1}{2})}_r$ generate the odd part of it. To describe the other sectors,
it is convenient to define
\bea
\label{redefineshs}
L^s_m &\equiv& \sqrt{ \frac{(N+s+1)!}{(2s+1)!(N-s)!} } T^s_m + \sqrt{ \frac{(N+s)!}{(2s-1)!(N-s-1)!} } U^s_m, \cr
V^s_m &\equiv& \sqrt{ \frac{(N+s+1)!}{(2s+1)!(N-s)!} } T^s_m - \sqrt{ \frac{(N+s)!}{(2s-1)!(N-s-1)!} } U^s_m, 
\eea
which generate $g^{(s)}$ and $h^{(s)}, s=3,4\ldots N-2$ respectively. Finally, $T^{N-1}_m$ generate $g^{(N-1)}$. 

In Appendix \ref{AppB}, we collect the structure constants, which can be derived via intertwining properties of the Clebsch-Gordan coefficients. Just as we have demonstrated for the $sl(3|2)$ case, one can invoke \eqref{structureconstants} to derive the supersymmetry transformation laws. In particular, we wish to write down the Killing spinor equation to classify the classical solutions. This relies on the commutation relations between $ \left\{ T^{s=0,1,2}_m, U^{s=0,1,2}_m \right\}$ and $\left\{ Q^{(\frac{1}{2})}_r, \bar{Q}^{(\frac{1}{2})}_r \right\}$, which we display below to be explicit. Consider first the commutation relations between the spin-0 and spin-1 fields with the $Q's$. Keeping the same notation as for the $sl(3|2)$ case, we denote the generators of the spin-1 multiplet $h^{(1)}$ by $A \equiv V^1_m$, we have\footnote{The commutation relations between $A$ and $\bar{Q}$ are identical, and thus omitted.}
\bea
\label{imptcomm}
[A_0, Q^{(\frac{1}{2})}_r ] &=& r \left( \frac{2N+1}{3}  \right)  Q^{(\frac{1}{2})}_r - \frac{\sqrt{2(N-1)(N+2)}}{3}  Q^{(\frac{3}{2})}_r,\cr \cr
[A_{\pm 1}, Q^{(\frac{1}{2})}_{\mp \frac{1}{2}} ] &=& 2 \sqrt{ \frac{(N+2)(N-1)}{6}  } Q^{(\frac{3}{2})}_{\mp \frac{3}{2}}, \cr \cr
[A_{\pm 1}, Q^{(\frac{1}{2})}_{\pm \frac{1}{2}} ] &=& \mp \left( \frac{2N+1}{3}    \right)  Q^{(\frac{1}{2})}_{\mp \frac{1}{2}} +   \frac{2}{3} \sqrt{ \frac{(N+2)(N-1)}{2}  } Q^{(\frac{3}{2})}_{\mp \frac{1}{2}}, \cr \cr
[U, Q^{(\frac{1}{2})}_r] &=& \frac{1}{\sqrt{N(N+1)}} Q^{(\frac{1}{2})}_r, \qquad [U, \bar{Q}^{(\frac{1}{2})}_r] = -\frac{1}{\sqrt{N(N+1)}} \bar{Q}^{(\frac{1}{2})}_r, 
\eea
noting that the gravitational $sl(2)$ subalgebra is generated by $L^1_m$. This serves as a consistency check that the $sl(2|1)$ sub-superalgebra is embedded in the same manner. Using \eqref{imptcomm}, we can write down the Killing spinor equation generalizing \eqref{spinor1} as
\be
\label{genspinor1}
\left(d + i\frac{1}{\sqrt{N(N+1)}} \cU - \frac{1}{2} \left( (e^a + \omega^a) + \left( \frac{2N+1}{3} \right) \Upsilon^a    \right)\gamma_a \right) \epsilon = 0.
\ee
We observe that it is essentially the same except for $N$-dependent scaling constants which can be absorbed into the various fields. The other fields do not play any role here, basically due to a simple constraint imposed by the Wigner-$6j$ symbols appearing in the structure constants in the relevant commutation relations. Recall that the Wigner-$6j$ symbol
\be
\label{6j}
\left\{\begin{array}{ccc}s & s' & s''\\ a & b & c \end{array} \right\} = 0\,\,\, \text{unless} \,\,\,s = |s' - s''|, \dots s+ s''.
\ee
For the supersymmetry transformation laws, the index $s'$ in \eqref{6j} takes the value of $\frac{1}{2}$. Since \eqref{genspinor1} is derived from the vanishing of $\delta \psi^{(\frac{1}{2})}$, the index $s''$ takes the value of $\frac{1}{2}$ as well, and the relevant values for $s$ are restricted to $\{0,1\}$. Thus, \eqref{genspinor1} takes on essentially the same form as \eqref{spinor1}. 

Similarly, for the analogues of \eqref{spinor2} in the $sl(N|N-1)$ higher-spin theory, for each $\frac{3}{2} \leq s'' \leq \frac{2N-3}{2}$, the vanishing of $\delta \psi^{(s'')}$ induces a constraint equation that involves even-graded fields of spin index $s=s'' \pm \frac{1}{2}$. Hence, the equation \eqref{spinor2} generalizes to a series of linear constraint equations (each labelled by $s$ and $r$) of the following form
\be
\label{genspinor2}
\sum_{l=-\frac{1}{2},\frac{1}{2}}\, \left(
\zeta^{(s-\frac{1}{2})}_{rl} \cV^{(s-\frac{1}{2})}_{l-r} + \zeta^{(s+\frac{1}{2})}_{rl} \cV^{(s+\frac{1}{2})}_{l-r} + \eta^{(s-\frac{1}{2})}_{rl} \cL^{(s-\frac{1}{2})}_{l-r} + \eta^{(s+\frac{1}{2})}_{rl} \cL^{(s+\frac{1}{2})}_{l-r} \right) \epsilon_l =0
\ee
which ensures the vanishing of the rest of the fermionic fields other than $\psi^{(\frac{1}{2})}$, i..e.
$$
\delta \psi^{(s)}_r = 0,\,\,\, s=\frac{3}{2}, \frac{5}{2}, \ldots \frac{2N-3}{2},\,\, |r| \leq s,
$$
where $\eta, \zeta$ are constant matrices that can be computed straightforwardly from the structure constants. We note that 
\be
\eta^{(1)} = 0,\qquad \zeta^{(N+\frac{1}{2})} = 0,
\ee
since $[L^1_m, \psi^{(\frac{1}{2})}] \sim \psi^{(\frac{1}{2})}$, and the multiplets $h^{(s)}$ terminate at $s = N-2$. Similar relations hold for the barred variables. 

We note that the above conclusions extend to the case of the infinite-dimensional algebra $shs[\lambda]$, since as explained in \cite{Fradkin:1990qk}, the generators \eqref{redefineshs} are those of $shs[\lambda]$, provided we analytically continue the integer $N$ to a positive real value $\lambda$ and abolish the restrictions to the index $s$. Further, there is a certain limit that involves taking $N\rightarrow \infty$, in which case we have a higher-spin gravity theory based on the infinite-dimensional algebra $shs[\infty]$.\footnote{As explained in \cite{Fradkin:1990qk}, this is the supersymmetric analogue of $hs[\infty]$ which can be physically understood as the algebra of area-preserving diffeomorphisms of 2D hyperboloids.} Essentially, we perform a redefinition of the generators as follows
\be
\label{redefinition}
\tilde{L}^s_m=N^{-s+1} L^s_m,\,\,
\tilde{V}^s_m=N^{-s} V^s_m,\,\,
\tilde{Q}^s_m = N^{1/2-s} Q^s_m,\,\,\tilde{\bar{Q}}^s_m = N^{1/2-s} \bar{Q}^s_m,
\ee
Then, we take the limit $N\rightarrow \infty$ and further perform another step of redefinition 
\be
L^s_m = \tilde{L}^s_m - \frac{(s-1)}{2} \tilde{V}^s_m,
\ee
noting that the gravitational $sl(2)$ is generated by $L^1_m$. In this limit, we can choose the generator $U_0$ to be normalized as
$$
U_0 = - L^0_0 - 2V^0_0
$$
and we check, using the new commutation relations obtained after this limiting procedure, that the Killing spinor equation \eqref{genspinor1} is preserved and reads
\be
\label{genspinor3}
\left(d + i\cU - \frac{1}{2} \left( (e^a + \omega^a) +  \frac{2}{3}  \Upsilon^a    \right)\gamma_a \right) \epsilon = 0.
\ee
Thus, the supersymmetry classification of ordinary solutions like the BTZ remains the same. Like in the finite-dimensional gauge algebra case, one can derive straightforwardly the rest of the supersymmetry transformation laws based on the structure constants - which we review in the Appendix \ref{AppB}.

\subsection{On solutions with higher spin fields}
\label{sec:onsol}
In the general $sl(N|N-1) \oplus sl(N|N-1)$ theory, the supersymmetry properties of solutions without higher-spin charges are thus identical as described earlier in Section \ref{sec:solvk}. In the following, we will briefly present the supersymmetric solutions with non-zero higher spin fields that we studied earlier in Section \ref{sec:holonomys} for the $sl(3|2)$ theory. 

We begin with the `highest-weight' ansatz 
\be
\label{genhighestweight}
\Gamma = b^{-1} \left( L_1 - \cL L_{-1} + \sum_{s=2}^{N-1} \cL^{(s)}_{-s} L^{(s)}_{-s} + \sum_{r=1}^{N-2} \cV^{(r)}_{-r} V^{(r)}_{-r} + \cU U_0 \right) b dx^+ + \Gamma_- dx^- + b^{-1}db, \,\,\,b \equiv e^{\rho L_0},
\ee
where we write
\be
\label{gammaminus}
\Gamma_- = \sum_{s=2}^{N-1} \sum_{m=-s}^{s} \mu^{(s)}_m L^{(s)}_m +  \sum_{r=1}^{N-2}  \sum_{m=-r}^{r} \Phi^{(r)}_m V^{(s)}_m.
\ee
Here we are interested in solutions of the form \eqref{genhighestweight} which preserve some amount of supersymmetry. Applying the constraints \eqref{genspinor2} in the $x^+$ direction, the higher-spin fields $\cL^{(s)}_{-s}$ and $V^{(r)}_{-r}$ in \eqref{genhighestweight} must vanish, upon which we can compute $\Gamma_-$ rather simply. The various generators transform under the embedded gravitational $sl(2)$ as
\be
\label{transformunder}
[L_m, V^{(s)}_r] = (sm - r) V^{(s)}_{m+r}, \qquad [L_m, L^{(s)}_r] = (sm - r) L^{(s)}_{m+r}.
\ee
Then, it is straightforward to show that for each multiplet in \eqref{gammaminus}, the various fields can be solved in terms of $\cL$ and $\{ \mu^{(s)}_s, \Phi^{(s)}_s \}$. Explicitly, we have
\bea
&&\mu^{(s)}_{s-2m} =  {}^sC_m \left( -\cL  \right)^m \mu^{(s)}_s,\,\,\, m=1,2,\ldots s, \\
&&\mu^{(s)}_{s-1} = \mu^{(s)}_{s-3} = \ldots \mu^{(s)}_{-s+1} = 0,
\eea
and identically for the fields $\Phi^{(s)}_r$. Applying the same constraint equations in the $x^-$ direction further kills off all fields, including $\cL$, except for $\mu^{(N-1)}_{N-1}$. Thus, the supersymmetric solution reads simply as 
\be
\label{finalsol}
\Gamma = \left( e^{\rho} L_1 + \cU U_0 \right) dx^+ + \left(  e^{(N-1)\rho} \mu^{(N-1)}_{N-1} L^{(N-1)}_{N-1}  \right) dx^- + L_0 d\rho,
\ee
with a similar expression for the anti-chiral sector. We observe that no component fields of the entire set of multiplets $h^{(s)}$ in \eqref{groupstructure} survive in a supersymmetric ansatz of the form \eqref{genhighestweight}. Apart from the highest-spin field $\mu^{(N-1)}_{N-1}$, this class of solutions is parametrized by the fields $\cU$ and $\tilde{\cU}$ which satisfy the quantization condition \eqref{Rchargeq}. Just like the $\cL=0$ solutions embedded in $osp(2|2)$ supergravity, in this case, we note that two real supercharges are preserved in each sector.


\subsection{On holonomy conditions and conical defects solutions}
\label{sec:onhol}
As discussed earlier, holonomy conditions play an important role in the study of $sl(N)\oplus sl(N)$ Chern-Simons theory. In Section \ref{sec:u1g}, we have briefly studied the $sl(3|2)$ case, and it is straightforward to state some results for the general case of $sl(N|N-1)$.  For global $AdS_3$, the gauge connection shares identical eigenvalues as $iL_0$, and the eigenvalues of $\oint \Gamma_\phi d\phi$ are those of $sl(N)$ and $sl(N-1)$. When exponentiated, the holonomy reads as
\be
\label{holi}
e^{\oint \Gamma_\phi d\phi} \sim  \left[ \begin{array}{c|c} \pm\mathbf{1}_{N\times N} & 0 \\ \hline
0 &\mp \mathbf{1}_{(N-1)\times (N-1)} \end{array}\right] ,
\ee
with the sign depending on whether $N$ is odd or even. We note that \eqref{holi} is, for any $N$, a linear combination of the identity and the $u(1)$ generator. We take both the supermatrix identity and \eqref{holi} to be the defining conditions for a smooth holonomy. We also note that it is straightforward to show that, apart from unimportant normalization constants which can be absorbed via a field redefinition, imposing the holonomy condition on \eqref{finalsol} yields also \eqref{Rchargeq}, which is also the (anti-)periodicity condition to be imposed on the Killing spinors. 

In the non-supersymmetric $sl(N)$ Chern-Simons theory, there is an interesting class of solutions which has been argued to be conical defects (and surpluses) spacetimes \cite{Castro:2011iw}. They play a critical role in the holographic duality conjecture, and we would like to investigate if there are natural generalizations of them in the supersymmetric theory. First, we begin with a brief review of some elementary aspects of these solutions following \cite{Castro:2011iw} but in the context of $sl(N|N-1) \oplus sl(N|N-1)$ theory.

Consider the ansatz
\be
\label{cdefecta}
\Gamma =   b^{-1}  \left( \sum_{k=1}^{2N-1} B_k(a_k,b_k) \right) b \,dx^+ + L_0 d\rho,\,\,\, \tilde{\Gamma} = - b \left( \sum_{k=1}^{2N-1} B_k(c_k,d_k) \right) b^{-1}\, dx^- - L_0 d\rho, b\equiv e^{\rho L_0},
\ee
with
$$
[ B_k (x,y) ]_{ij} = x \delta_{i,k} \delta_{j,k+1} - y \delta_{i,k+1} \delta_{j,k}.
$$
The constant matrices $B_k$ are linear combination of weight-one generators (i.e. $T_{\pm 1}, U_{\pm 1}$). They are diagonalizable with imaginary eigenvalues. The interesting solutions are found by imposing three essential conditions, namely (i)the metric induced by the connection is locally $AdS_3$, (ii)the holonomy along the $\phi$-direction is trivial and (iii)the stress-energy tensor is negative and bounded from below by its value for global $AdS_3$. 

To conveniently list the equivalence classes of solutions, we can restrict the ansatz to consist of the following maximal commuting set as parametrized by $B_k$.
\bea
&&\text{For even}\,\,N,\, k=1,3,\dots, 2N-3. \cr
&&\text{For odd}\,\,N,\, k=1,3,\dots N-2, N+1,\dots, 2N-2.
\eea
with $a_k=b_k=c_k=d_k$ for all cases. Our choice is slightly different from the $sl(N) \oplus sl(N)$ theories considered in \cite{Castro:2011iw} for the cases of odd $N$, due to the fact that we are now taking a supertrace instead of the ordinary matrix trace. Then, from \eqref{cdefecta}, we find the line element
\be
\label{ads3metric}
ds^2 = d\rho^2 - \frac{1}{\text{str} \left( L_0^2 \right)} \left(  e^{\rho} + \Lambda e^{-\rho} \right)^2 dt^2
+ \frac{1}{\text{str} \left( L_0^2 \right)} \left(  e^{\rho} - \Lambda e^{-\rho} \right)^2 d\phi^2,
\ee 
\be
\label{abtLambda}
\Lambda \equiv \frac{1}{2} \left( \sum_{k=1}^{\lfloor  \frac{N}{2} \rfloor } a_{2k-1}^2 - \sum_{k=1}^{N-1-\lfloor \frac{N}{2} \rfloor} a^2_{2k+N-1} \right) = -\frac{1}{4}\text{str} \left(  \Gamma_+^2  \right),
\ee
where we have shifted $\rho \rightarrow \rho + \text{log} (\sqrt{\Lambda}) $. Note that the negative sign in \eqref{abtLambda} is due to the supertrace convention. The metric is locally $AdS_3$ and we can interpret the higher-spin fields as topological matter, with $\Lambda$ capturing its global properties. Since $\delta \phi =2\pi$, one can compute the conical deficit $\delta_c$ of the spacetime \eqref{ads3metric} to be 
\be
\label{conicaldefect}
\delta_c = 2\pi \left( 1 - \sqrt{\frac{4\Lambda}{\text{str}(L_0^2)}} \right),\qquad \text{str} ( L_0^2 ) = \frac{N(N-1)}{4}.
\ee
Further, the stress tensor for this class of solutions (in units of $l_{AdS}/G$) can be expressed as 
$$
M=-\frac{8k\Lambda}{N(N-1)}.
$$
In the case of global $AdS_3$, the gauge connection $\Gamma_+$ has identical eigenvalues as that of the complex generator $iL_0$, and a similarity transformation brings it to the form above, with $\Lambda=\frac{1}{4} \text{str} ( L_0^2 )$, yielding $\delta_c=0, M=-k/2$ expectedly. This yields an upper bound for $\Lambda$ as set by global $AdS_3$ (and thus lower bound for the mass $M$) for conical defect spacetimes
\be
\label{bound1}
0<\Lambda < \frac{N(N-1)}{16}.
\ee
Using both the holonomy condition \eqref{holi} and the trivial one, we can determine the set of parameters $\{a_k\}$ that defines this class of spacetime. As the simplest example, for the $sl(3|2) \oplus sl(3|2)$ theory, the holonomy condition \eqref{holi} translates into requiring $a_1$ to be an integer and $a_4$ to be a half-integer, whereas the trivial holonomy condition imposes both parameters to be integral. The valid interval \eqref{bound1} reads as
\be
\label{upperboundsum}
0 < a_1^2 - a_4^2 < \frac{3}{4},
\ee
where the possible values of the $2\Lambda$ are thus just either $1/4$ or $1/2$. These Diophantine equations can be solved to show that for both cases, there are no solutions. Thus, analogous to the $sl(3)\oplus sl(3)$ case, the $sl(3|2)\oplus sl(3|2)$ theory contains no conical defect spacetimes. 

Let us move on to consider the $sl(4|3) \oplus sl(4|3)$ theory for which the bound \eqref{bound1} reads as
\be
\label{upperboundsum2}
0 < a_1^2 + a_3^2 - a_5^2 < \frac{3}{2},
\ee
with the holonomy conditions translating into solutions of trivariate Diophantine equations:
\newline
\newline
(i)For the holonomy \eqref{holi}, both $a_1, a_3$ half-integers and $a_5$ is integral, with all possible solutions taking the mass $M=-\frac{k}{6}$. There is an infinite set of solutions. An example for which it is easy to obtain a closed form is the subset defined by $a_1=a_3\equiv 2y+1$ which reduces to a Pell-like equation which we find to have the solution $8y = -4+2((3-2\sqrt{2})^r+(3+2\sqrt{2})^r), r\in \mathbb{Z}^+$. 
\newline
\newline
(ii)For the trivial holonomy,  all parameters are integral with the mass $M=-\frac{k}{3}$. There is also an infinite set of solutions. A rather simple example is the family of solutions defined by $a_1 = a_5, a_3=1$. 
\newline
\newline
Thus, the solution space is notably distinct from the $sl(4) \oplus sl(4)$ theory, where there are only three distinct solutions labelled by $(a_1, a_3)$ as follows: $\left(\frac{1}{2}, \frac{1}{2} \right), \left(1, 0 \right)$ and  $\left(1, 1 \right)$. One can attribute the infinitely-degenerate spectrum in the case of the $sl(4|3) \oplus sl(4|3)$ theory to the fact that $\Lambda$ is the difference between two separate sets of $a_k^2$ (see \eqref{abtLambda}) in this case as we are computing supertraces. But in the $sl(4) \oplus sl(4)$ case, $\Lambda$ is simply the sum of all $a_k^2$ and thus for any given finite mass, the degeneracy is finite. It is straightforward to carry out a similar analysis for other values of $N$. These solutions generically break all supersymmetries. It would be interesting to understand these solutions from the holographic CFT point of view, just as in the case of the $sl(4)$ theory in \cite{Castro:2011iw}. This would give us a clearer picture of their physical significance and interpretation. We leave these investigations to future work.

\section{Discussion}\label{sec:discussion}

In this paper, we have initiated a study of some interesting aspects of $\cN=(2,2)$ higher-spin SUGRA that is defined on direct sum of two copies of finite-dimensional $sl(N|N-1)$ gauge algebras, using the case of $N=3$ as our prime example. We derive explicitly the supersymmetry transformation laws, and in particular, noted that global $AdS_3$ preserves eight supercharges, the massless BTZ preserving four and the extremal BTZ preserving two. By performing an analytic continuation on $N$, we have also checked that this classification holds in the infinite-dimensional $shs[\lambda]$ gauge algebra case.

We have briefly discussed a class of supersymmetric solutions (preserving four supercharges) which are massless BTZ solutions carrying higher-spin fields (they are actually the RG flow solutions of \cite{Kraus2} carrying some $U(1)$ charge). We have constructed a class of higher-spin black hole solutions which are natural generalizations of those in the $sl(N)$ theories, and which, like the latter, establish a remarkable relation between the BTZ holonomy condition and integrability of higher-spin charges. Also, we find that there are infinitely many smooth conical defect solutions instead of a finite number of them in any particular $sl(N|N-1)$ theory for $N\geq 4$. 

The consistency of some aspects of our computations was verified by checking that, upon truncating some of the gauge fields, we obtain ordinary SUGRA defined via Chern-Simons with $osp(2|2)$ and $osp(1|2)$ gauge algebras which are sub-superalgebras of $sl(N|N-1)$. From a partial analysis of the asymptotic symmetry algebra, we recovered the $\cN=2$ superconformal algebra (for each chiral sector), and computed explicitly the Sugawara redefinition of the energy-momentum tensor. The $u(1)$ field generates spectral flows of bulk solutions, and is quantized to ensure either the $\phi$-periodicity of the Killing spinor or smooth holonomies. 

Let us end off by briefly indicating some suggestions for future work. Generally, it would be important to develop our understanding of the holographic duality by deriving in detail what aspects of the boundary CFT can be related to the physics in the bulk. For example, in the non-supersymmetric case, the free energy of the higher-spin black holes in $hs[\lambda] \oplus hs[\lambda]$ Chern-Simons theory has been successfully reproduced from the dual CFT side in \cite{Gaberdiel:2012yb} and \cite{Kraus3}, relying on the fact that higher-spin corrections can be computed at high temperature from correlation functions of $\cW-$algebra currents. It would be interesting to see how this works out for the cases where the gauge algebra is $shs[\lambda] \oplus shs[\lambda]$. 

Another natural avenue for further study is to investigate the properties, both from the bulk and boundary perspectives, of the higher-spin black holes and conical defect spacetimes discussed in Sections \ref{sec:four} and \ref{sec:five}. For example, in the arena of higher-spin black hole thermodynamics, it was recently shown in \cite{Banados:2012ue}\footnote{See also \cite{Perez:2012cf} for a clarifying discussion.} that the relation between integrability conditions and smooth holonomy condition can be understood fundamentally as arising from a careful computation of the on-shell Chern-Simons action. It may be interesting to re-visit this computation in the context of $sl(N|N-1)$ theories.


\section*{Acknowledgments}
I am indebted to Marc Henneaux, Per Kraus, an anonymous referee, and in particular, Ori Ganor, for various advice on a previous draft. I also thank Emanuele Latini and Andrew Waldron for sharing with me some of their insights on higher-spin gravity. I acknowledge financial support drawn from the Victor Lenzen Memorial fund of BCTP during the course of completion of this work.

\renewcommand{\theequation}{A-\arabic{equation}}  
\setcounter{equation}{0}  

\begin{appendix}

\section{On the isomorphism $sl(2|1) \simeq osp(2|2)$ }
\label{AppendixA}
We begin by displaying the commutation relations for $osp(2|2)$ as reviewed in the Appendix of \cite{Henneaux:2012ny}. $R^{\pm}_i, i=1,2$ are the four fermionic generators; $E,F,H$ generate the $sl(2)$ while $J_{12}=-J_{21}$ is the $u(1)$ generator. 
\bea
&&i\{ R_i^+, R_j^- \} = J_{ij} - \delta_{ij}\,H,\qquad i\{ R_i^-, R_j^- \} = - 2\delta_{ij}\,F,\qquad i\{ R_i^+, R_j^+ \} =  2\delta_{ij}\,E. \cr \cr
&&[H, R_i^+] = R_i^+,\,\,\, [F,R_i^+]=R_i^-,\,\,\,[H,R_i^-]=-R_i^-,\,\,\,[E,R_i^-]=R_i^+,\,\,\,[F,R_i^-]=0. \cr \cr
\label{osp}
&&[H,E]=2E,\,\,\,[H,F]=-2F,\,\,\,[E,F]=H,\,\,\, [J_{ij}, R_k^{\pm}]= \delta_{jk} R_i^{\pm} - \delta_{ik} R_j^{\pm}
\eea
We verify that this algebra is isomorphic to $sl(2|1)$ via the following identifications between the generators of the latter (written in our choice of basis) and those  in \eqref{osp} :
\bea
&&E \sim L_-,\,\,\, F \sim -L_+,\,\,\, H \sim 2L_0, \cr \cr
\label{sl32}
&&J_{12} \sim i6U_0,\,\,\, Q_{\pm \frac{1}{2}} \sim \frac{-i}{2\sqrt{3}} \left( R_1^{\pm} + i R_2^{\pm}  \right),\,\,\,\bar{Q}_{\pm \frac{1}{2}} \sim \frac{1}{2\sqrt{3}} \left( R_1^{\pm} - i R_2^{\pm}  \right),
\eea
We note from \eqref{sl32} that the matching between $sl(2|1)$ and $osp(2|2)$ involves changing the reality conditions of the generators $U_0$ and $Q_{\pm \frac{1}{2}}$. In our paper, the classification of solutions based on the number of real supersymmetries preserved is performed after continuing $U_0 \rightarrow iU_0$ and treating $\psi,\bar{\psi}$ as complex Grassmann variables. Although $sl(2|1)$ and $osp(2|2)$ are isomorphic algebras,  at the level of representation, $sl(2|1)$ and $osp(2|2)$ are distinct. The latter admits only real representations whereas the former admits both real and complex ones. 

\renewcommand{\theequation}{B-\arabic{equation}}  
\setcounter{equation}{0}  

\section{Variations of the $osp(2|2)$ gauge fields}
\label{AppA}
In Section \ref{sec:recover}, we recovered the $\cN=2$ superconformal algebra by considering the variations of the fields $\varphi_+, \bar{\varphi}_+, \cL$ and  $\cU$  under a gauge transformation that preserves the form of the highest-weight ansatz. Although we do not need the full expressions for these variations in our computation, we collect them here for completeness and verification. 
\bea
\label{111}
\delta \varphi_+ &=& \frac{3}{2} \varphi_+ \xi'\, + \varphi_+' \xi \, + \frac{1}{6}\cU \varphi_+ \xi + \frac{5}{6} \varphi_+ \alpha' + \sqrt{\frac{8}{3}} \Psi_+ \alpha 
- \frac{1}{6}\varphi_+ \eta + \left( \frac{5}{3} \Upsilon + \cL + \frac{1}{6} \cU' +\frac{1}{36}\cU^2  \right) \nu_-\cr
&&+ \frac{1}{3} \cU \nu_-' + \nu_-''+5\sqrt{\frac{2}{3}} \Psi_+ \chi' + \left( 4\sqrt{\frac{2}{3}} \Psi_+' + \frac{16}{9} \Upsilon \varphi_+ + \sqrt{\frac{8}{27}} \cU \Psi_+  \right) \chi  \cr
&&+\left( -\sqrt{\frac{8}{3}} \Upsilon' + 4\sqrt{\frac{2}{3}} \cW - \frac{4}{9} \sqrt{\frac{2}{3}}\Upsilon \cU   \right) \zeta_- - \left( \frac{8}{3}\sqrt{\frac{2}{3}} \Upsilon   \right) \zeta_-'  \cr
\label{222}
\delta \bar{\varphi}_+ &=& \frac{3}{2} \bar{\varphi}_+ \xi'\, + \bar{\varphi}_+' \xi \, - \frac{1}{6}\cU \bar{\varphi}_+ \xi + \frac{5}{6} \bar{\varphi}_+ \alpha' - \sqrt{\frac{8}{3}} \bar{\Psi}_+ \alpha  + \frac{1}{6}  \bar{\varphi}_+ \eta   + \left( \frac{5}{3} \Upsilon + \cL + \frac{1}{6} \cU' +\frac{1}{36}\cU^2  \right) \bar{\nu}_-   \cr
&& + \frac{1}{3} \cU \bar{\nu}_-' + \bar{\nu}_-'' +5\sqrt{\frac{2}{3}} \bar{\Psi}_+ \chi' + \left( 4\sqrt{\frac{2}{3}} \bar{\Psi}_+' - \frac{16}{9} \Upsilon \bar{\varphi}_+ - \sqrt{\frac{8}{27}} \cU \bar{\Psi}_+  \right) \chi \cr
&&+ \left( \sqrt{\frac{8}{3}} \Upsilon' + 4\sqrt{\frac{2}{3}} \cW - \frac{4}{9} \sqrt{\frac{2}{3}}\Upsilon \cU   \right) \bar{\zeta}_- + 
\left( \frac{8}{3}\sqrt{\frac{2}{3}} \Upsilon   \right) \bar{\zeta}_-'   \\ \cr 
\label{333}
\delta \cU &=& \eta' - \bar{\Psi}_+ \zeta_{-} + \Psi_+ \bar{\zeta}_{-} + \bar{\varphi}_+ \nu_{-} - \varphi_+ \bar{\nu}_{-} \\ \cr
\label{444}
\delta \cL &=& \frac{1}{2}\xi''' + 2\cL \xi' + \cL' \xi + \Upsilon' \alpha + 2\Upsilon \alpha' + \left(  
\frac{17}{12}\sqrt{\frac{2}{3}}\left(  \Psi_+ \bar{\varphi}_+ + \bar{\Psi}_+ \varphi_+ \right)\chi - 4\cW' - \frac{5}{6}\left( \bar{\varphi}_+ \varphi_+ \right)' \right)  \chi \cr
&&+\left(  \frac{5}{36} \left( \varphi_+ \bar{\varphi}_+ \right)  - 6\cW \right) \chi' 
- \left(  \frac{1}{18}\cU \bar{\varphi}_+ + \frac{1}{6} \bar{\varphi}'_+ + \frac{5}{4\sqrt{6}}\bar{\Psi}_+ \right)\nu_- - \frac{1}{2}\bar{\varphi}_+ \nu'_- \cr
&&+\left(  \frac{5}{8} \sqrt{\frac{3}{2}} \left( \Upsilon - \cL  \right)  + \frac{5\cU'}{144\sqrt{6}} - \frac{5\cU^2}{864\sqrt{6}}\right)\bar{\varphi}_+ \zeta_- + \left( \frac{5}{8\sqrt{6}} \bar{\varphi}_+'' + \frac{5\cU}{72\sqrt{6}} \bar{\varphi}_+' - \frac{1}{8} \bar{\Psi}_+'  -\frac{1}{72} \cU  \bar{\Psi}_+  \right)\zeta_-' \cr
&&-\left(  \frac{5}{6\sqrt{6}} \bar{\varphi}_+' + \frac{5}{24}\bar{\Psi}_+  \right) \zeta_-' - \frac{5\bar{\varphi}_+}{12\sqrt{6}} \zeta_-''  +\left(  \frac{5}{8} \sqrt{\frac{3}{2}} \left(-\Upsilon + \cL  \right)  + \frac{5\cU'}{144\sqrt{6}} + \frac{5U_0^2}{864\sqrt{6}}\right)\varphi_+ \bar{\zeta}_- \cr
&& + \left( \frac{5}{8\sqrt{6}} \varphi_+'' + \frac{5U_0}{72\sqrt{6}} \varphi_+' - \frac{1}{8} \Psi_+'  +\frac{1}{72} \cU  \Psi_+  \right) \bar{\zeta}_-' +\left(  \frac{5}{6\sqrt{6}} \varphi_+' - \frac{5}{24}\Psi_+  \right) \bar{\zeta}_-' + \frac{5\varphi_+}{12\sqrt{6}} \bar{\zeta}_-'' \\ \cr
\delta \Upsilon &=& \Upsilon' \xi + 2\Upsilon \xi' +  2\cL \alpha' + \cL'\alpha + \frac{1}{2}\alpha'' +\sqrt{\frac{3}{32}}\left( \bar{\Psi}_+ \nu_- - \Psi_+ \bar{\nu}_- \right) \cr 
&&+ \left(   \frac{5}{4} \varphi_+ \bar{\varphi}_+ - 6\cW \right) \chi' + \left( \frac{5}{6} \left( \varphi_+ \bar{\varphi}_+  \right)' +\frac{25}{24} \left( \bar{\varphi}_+ \Psi_+ + \varphi_+\bar{\Psi}_+  \right) - 4\cW'  \right)\chi \cr
&&+ \left( \sqrt{\frac{3}{128}} \bar{\varphi}_+'' + \frac{\cU}{24\sqrt{6}}\bar{\varphi}_+' + 
\left(  \frac{\cU'}{16\sqrt{6}}+\sqrt{\frac{27}{128}}(\cL-\Upsilon) + \frac{\cU^2}{288\sqrt{6}}                \right)\bar{\varphi}_+ +\frac{3}{4} \bar{\Psi}_+' + \frac{\cU}{24} \bar{\Psi}_+  \right)\zeta_- \cr
&&+\frac{5}{8\sqrt{6}} \bar{\varphi}_+ \zeta_-'' + \left( \bar{\Psi}_+ + \frac{1}{\sqrt{6}} \bar{\varphi}_+' + \frac{\cU}{12\sqrt{6}}\bar{\varphi}_+   \right) \zeta_-'
\eea
where the superscripted primes refer to derivatives with respect to $\phi$.

\renewcommand{\theequation}{C-\arabic{equation}}  
\setcounter{equation}{0}  

\section{Structure constants}
\label{AppB}
Following Racah \cite{Fradkin:1990qk}, introduce a basis as follows, where $E_{ij}$ refers to a matrix with unity in the $i^{\text{th}}$ row and  $j^{\text{th}}$ column (our convention for the even-graded generators differs slightly from \cite{Fradkin:1990qk}, but is presented in full here for clarity)
\bea
T^s_m &=& \sqrt{ \frac{2s+1}{N+1}} \sum_{r,q} C^{\frac{N}{2} s \frac{N}{2}}_{r\,m\,q} E_{\frac{N}{2} + 1 - q, \frac{N}{2} + 1 - r},\,\,\,(s=0,1, \ldots N) \cr
U^s_m &=& \sqrt{ \frac{2s+1}{N}} \sum_{r,q} C^{\frac{N-1}{2} s \frac{N-1}{2}}_{r\,\,\,m\,\,\,q} E_{\frac{3(N+1)}{2} - q, \frac{3(N+1)}{2} + 1 - r},\,\,\,(s=0,1, \ldots N-1)  \cr
Q^s_m &=& \sqrt{ \frac{2s+1}{N}} \sum_{r,q} C^{\frac{N}{2} s \frac{N-1}{2}}_{r\,m\,q} E_{\frac{3(N+1)}{2} - q, \frac{N}{2} + 1 - r},\,\,\,(s=\frac{1}{2},\frac{3}{2}, \ldots N-\frac{1}{2}) \cr
\bar{Q}^s_m &=& \sqrt{ \frac{2s+1}{N+1}} \sum_{r,q} C^{\frac{N-1}{2} s \frac{N}{2}}_{r\,m\,q} E_{\frac{N}{2}+1-q, \frac{3(N+1)}{2} - r},\,\,\,(s=\frac{1}{2},\frac{3}{2}, \ldots N-\frac{1}{2}) 
\eea
The structure constants read
\bea
\label{structureconstants}
[T^s_m, Q^{s'}_{m'}] &=& \sum_{s'', m''} \left( -1 \right)^{(s+s'-\frac{3}{2}+N+m)} \sqrt{(2s+1)(2s'+1)} \left\{\begin{array}{ccc}s & s' & s''\\ \frac{N-1}{2} & \frac{N}{2} & \frac{N}{2} \end{array} \right\} C^{s\,s'\,s''}_{-m\,m'\,m''}  Q^{s''}_{m''} \cr \cr
[T^s_m, \bar{Q}^{s'}_{m'}] &=& \sum_{s'', m''} \left( -1 \right)^{(s''-\frac{1}{2}+N+m)} \sqrt{(2s+1)(2s'+1)} \left\{\begin{array}{ccc}s & s' & s''\\ \frac{N-1}{2} & \frac{N}{2} & \frac{N}{2} \end{array} \right\} C^{s\,s'\,s''}_{-m\,m'\,m''}  \bar{Q}^{s''}_{m''} \cr\cr
[U^s_m, Q^{s'}_{m'}] &=& \sum_{s'', m''} \left( -1 \right)^{(2(s+s')-\frac{1}{2}+N+m-s'')} \sqrt{(2s+1)(2s'+1)} \left\{\begin{array}{ccc}s & s' & s''\\ \frac{N}{2} & \frac{N-1}{2} & \frac{N-1}{2} \end{array} \right\} C^{s\,s'\,s''}_{-m\,m'\,m''}  Q^{s''}_{m''} \cr \cr
[U^s_m, \bar{Q}^{s'}_{m'}] &=& \sum_{s'', m''} \left( -1 \right)^{(s+s'+\frac{1}{2}+N+m)} \sqrt{(2s+1)(2s'+1)} \left\{\begin{array}{ccc}s & s' & s''\\ \frac{N}{2} & \frac{N-1}{2} & \frac{N-1}{2} \end{array} \right\} C^{s\,s'\,s''}_{-m\,m'\,m''}  \bar{Q}^{s''}_{m''} \cr \cr
[T^s_m, T^{s'}_{m'}] &=& \sum_{s'', m''} \left( -1 \right)^{(s''+N)} \left(1-(-1)^{s-s'-s''}  \right) \sqrt{(2s+1)(2s'+1)} \left\{\begin{array}{ccc}s & s' & s''\\ \frac{N}{2} & \frac{N}{2} & \frac{N}{2} \end{array} \right\} C^{s\,s'\,s''}_{m\,m'\,m''}  T^{s''}_{m''} \cr \cr
[U^s_m, U^{s'}_{m'}] &=& -\sum_{s'', m''}  (-1)^{(s''+N)} \left(1-(-1)^{s-s'-s''}  \right) \sqrt{(2s+1)(2s'+1)} \left\{\begin{array}{ccc}s & s' & s''\\ \frac{N-1}{2} & \frac{N-1}{2} & \frac{N-1}{2} \end{array} \right\} \cr
&&\qquad \qquad \times C^{s\,s'\,s''}_{m\,m'\,m''}  U^{s''}_{m''} \cr \cr
\left\{ Q^s_m, \bar{Q}^{s'}_{m'} \right\} &=& \sum_{s'', m''} C^{s\,s'\,s''}_{-m\,m'\,m''} \sqrt{(2s+1)(2s'+1)} \Bigg(
\left( -1 \right)^{(s+s'-1+N)} \left\{\begin{array}{ccc}s & s' & s''\\ \frac{N}{2} & \frac{N}{2} & \frac{N-1}{2} \end{array} \right\}  T^{s''}_{-m''} \cr 
&&\qquad \qquad+ \left( -1 \right)^{(s''+N)}  \left\{\begin{array}{ccc}s & s' & s''\\ \frac{N-1}{2} & \frac{N-1}{2} & \frac{N}{2} \end{array} \right\}  U^{s''}_{-m''} \Bigg) 
\eea
The above formulae are used in explicit computations in the $sl(3|2)$ Chern-Simons theory in Section \ref{sec:four}. Please note that $\left\{\begin{array}{ccc}s & s' & s''\\ a & b & c \end{array} \right\}$ are the Wigner $6j$ symbols, while $C^{s\,s'\,s''}_{m\,m'\,m''} $ denote Clebsch-Gordan coefficients.

\end{appendix}

\end{document}